\documentclass[aps,prx,amsmath,floats,floatfix,twocolumn,
  superscriptaddress,nofootinbib,showpacs]{revtex4-1}

\pdfoutput=1 % Specifiy PDFLaTeX as per arXiv request. This has to be in the first 5 lines!
\usepackage{diagbox}
\usepackage{mathrsfs}
\usepackage{multirow}
\usepackage{braket}
\usepackage{amsfonts}
\usepackage{amsmath}
\usepackage{amssymb}
\usepackage{amsthm}
\usepackage{bm}
\usepackage{graphicx}
\usepackage{graphics}
\usepackage{latexsym}
\usepackage{rotating}
\usepackage[dvipsnames]{xcolor}
\usepackage{mathrsfs}
\usepackage{microtype}
\usepackage{verbatim}
\usepackage{url}
\usepackage{yfonts}
\usepackage{float}
\usepackage{xspace}
\usepackage{mathrsfs}
\usepackage{siunitx}
\usepackage{array}
\usepackage{dashrule}
\usepackage[T1]{fontenc}
\usepackage{phonetic}
\usepackage[normalem]{ulem}

\usepackage{tikz}
\usetikzlibrary{arrows.meta,positioning,calc}

\usepackage[breaklinks=true]{hyperref}
\hypersetup{
    colorlinks=true,
    linkcolor=NavyBlue,
    filecolor=Magenta,
    urlcolor=NavyBlue,
    citecolor=NavyBlue
}

\allowdisplaybreaks[4]

\def\pheq{\phantom{=}}
\newcommand{\half}{\dfrac{1}{2}}

\renewcommand{\thorn}{\mathord{\text{\fontseries{l}\selectfont\th}}}
\renewcommand{\eth}{\mathord{\text{\fontseries{l}\selectfont\dh}}}

\makeatletter
\newcommand*{\rom}[1]{\expandafter\@slowromancap\romannumeral #1@}
\makeatother

\allowdisplaybreaks

\begin{document}

\title{Einstein Reconstruction Scheme: \\ Metric Reconstruction by Direct Inversion of the Einstein Operator}

\author{Kelvin~Ka-Ho~Lam} 
\email{khlam4@illinois.edu}
\affiliation{Illinois Center for Advanced Studies of the Universe \& Department of Physics, University of Illinois Urbana-Champaign, Urbana, Illinois 61801, USA}

\author{Dongjun Li}
\email{dongjun@illinois.edu}
\affiliation{Illinois Center for Advanced Studies of the Universe \& Department of Physics, University of Illinois Urbana-Champaign, Urbana, Illinois 61801, USA}

\author{Nicol\'as Yunes}
\email{nyunes@illinois.edu}
\affiliation{Illinois Center for Advanced Studies of the Universe \& Department of Physics, University of Illinois Urbana-Champaign, Urbana, Illinois 61801, USA}

\date{\today}

%%%%%%%%%%%%%%%%%%%%%%%%%%%%%%%%%%%%%%%%%%%%%%%%%%%%%%%%%%%%%%
\begin{abstract} 

%Abstract limited to 600 characters!!

We propose a metric reconstruction scheme that computes the metric directly from the linearized Einstein equations for noncompact sources. 
Working in a traceful radiation gauge, we hierarchically solve four Einstein transport equations, with the remaining six enforced by suitable homogeneous data. 
Unlike previous Newman-Penrose approaches, we show how the homogeneous data are fixed, thereby closing the full set of Einstein equations. 
We conclude by relating our scheme to existing sourced reconstruction methods, thus establishing a connection between the Newman-Penrose and Hertz-potential approaches.

\end{abstract}

%\keywords{Suggested keywords}%Use showkeys class option if keyword display desired
\maketitle

\vspace{0.2cm}
\noindent 
\textit{Introduction.}\textemdash \
The perturbation $h_{\mu\nu}$ of a background spacetime geometry is governed by the linearized Einstein equations with some external source $T_{\mu\nu}$. 
Such metric perturbations, therefore, allow us to investigate the stability of the background spacetime \cite{Dafermos:2016uzj, Andersson:2019dwi} and the dynamics induced by matter sources. 
However, the linearized Einstein equations form a coupled system of partial differential equations for which we lack systematic methods to find analytic solutions in backgrounds without enhanced symmetries, such as spherical symmetry \cite{Regge:1957td, Zerilli:1970se}.

Instead of working with metric perturbations, one could work with perturbations of the curvature directly. In Petrov type-D spacetimes\footnote{For simplicity, we restrict ourselves to vacuum backgrounds of Petrov type-D;
nevertheless, the Einstein reconstruction procedure we develop extends to Petrov type-II backgrounds if one uses the radiation gauge in which $h_{\mu\nu}$ remains transverse to the repeated principal null direction.}, one can derive the Teukolsky equation, a separable partial differential equation for the perturbed Weyl scalars $\Psi_0$ or $\Psi_4$, which encodes the radiative information of $h_{\mu\nu}$ for dynamical perturbations \cite{Teukolsky:1973ha, Teukolsky:1974yv}, or tidal information for static deformations \cite{Hu:2026ute, Li:2026hpw}. 
The solution to the Teukolsky equation for the perturbed Weyl scalars can be readily found using frequency and angular mode decomposition methods \cite{Teukolsky:1973ha, Teukolsky:1974yv}. 
In first-order perturbation theory, if one is only concerned with the perturbation's radiative information, the Weyl scalars suffice to compute many observables, such as gravitational fluxes at infinity and at the event horizon \cite{Pound:2021qin, Teukolsky:1974yv, Chandrasekhar:1985kt}. 
However, these perturbed scalars do not fully encode matter-induced static perturbations and zero-mode shifts, and thus, extracting complete information requires the full $h_{\mu\nu}$.
Moreover, higher-order perturbation theory requires the full $h_{\mu\nu}$ to compute the stress-energy tensor sourced by products of lower-order perturbations.
Performing second-order perturbation theory is important for accurate modeling of extreme-mass-ratio inspirals (EMRIs) \cite{Amaro-Seoane:2014ela, Barack:2018yly, Pound:2021qin, Upton:2021oxf, Upton:2023tcv} and possibly the ringdown stage of a binary black hole coalescence \cite{Ma:2024qcv, Yi:2024elj, Lagos:2024ekd, Khera:2024bjs, Khera:2026lna}, both in vacuum and in astrophysical environments \cite{Dyson:2025dlj, Li:2025ffh, Li:2026foy}, for next-generation gravitational-wave detectors.

Various approaches have been developed to reconstruct the remaining metric components from the perturbed Weyl scalars. One of the most widely used reconstruction schemes is the Chrzanowski-Cohen-Kegeles (CCK) approach \cite{Cohen:1975qjg, Chrzanowski:1975wv, Kegeles:1979an, Wald:1978vm}, in which $h_{\mu\nu}$ in a radiation gauge is obtained via two steps. 
First, one solves a fourth-order differential equation for a Hertz potential $\Phi$, sourced by Weyl scalars \cite{Lousto:2002em, Ori:2002uv}, with the help of the Teukolsky-Starobinsky identities \cite{Teukolsky:1974yv, Starobinskii:1974nkd, Berens:2024czo}. 
Then, $h_{\mu\nu}$ is found by acting with a second-order differential operator on $\Phi$. 
While the CCK approach is useful for vacuum perturbations, it does not generically apply in the presence of sources, and $\Phi$ does not contain information on the zero-mode perturbations. 
Recently, Green-Hollands-Zimmerman (GHZ) developed an extension to the CCK approach by including a corrector tensor that satisfies second-order transport equations along a principal null direction \cite{Green:2019nam, Hollands:2024iqp, Casals:2026man}. 
This captures the additional metric perturbations that could not be reconstructed by the CCK approach. 
However, the GHZ formulation also presents several challenges: (i) static perturbations and zero modes must be treated separately, and (ii) trivial initial/boundary data are often assumed, which might be incompatible with extended sources with weaker decay properties. 

Alternatively, one can employ a Newman-Penrose (NP) reconstruction scheme, which hierarchically computes $h_{\mu\nu}$ in radiation gauges by solving NP transport equations along a principal null direction. 
This approach was developed in \cite{Chandrasekhar:1985kt, Loutrel:2020wbw, Ripley:2020xby, Suvorov:2019qow}, and was recently extended by \cite{Li:2026rkf} to include generic $T_{\mu\nu}$. 
In this method, static modes induced by $T_{\mu\nu}$ can be computed alongside the radiative modes.
Moreover, provided that the initial/boundary data $h_{\mu\nu}|_{\Sigma}$ satisfies a subset of Einstein equations on a hypersurface $\Sigma$ transverse to the principal null direction, one can show that $h_{\mu\nu}$ satisfies all Einstein equations throughout the transport domain \cite{Li:2026rkf}. 
In spite the potential of this method, nobody has yet shown whether such $h_{\mu\nu}|_{\Sigma}$ exists, and if it does, how one would find it in concrete applications.

In this Letter, we introduce another reconstruction approach, which we refer to as the Einstein reconstruction scheme. 
This approach works directly with the Einstein operator in Geroch-Held-Penrose (GHP) form and in a class of traceful radiation gauges \cite{Price:2007dev, Green:2019nam, Toomani:2021jlo, Hollands:2024iqp, Spiers:2023cip, Casals:2026man}.
We show that the hierarchical reconstruction procedure is not restricted to the NP transport system, but rather, it also applies to the Einstein equations, the underlying system of other sourced metric reconstruction frameworks \cite{Aksteiner:2016pjt, Green:2019nam, Wardell:2024yoi}.
We invert four components of the linearized Einstein equations by solving second-order transport equations using Held's integration technique \cite{Held:1974for, Held:1976kil}.
Using the Wald \cite{Wald:1978vm} and Bianchi identities, the remaining Einstein equations can be satisfied by choosing suitable homogeneous solutions for the transport equations, thus showing that a unique $h_{\mu\nu}|_{\Sigma}$ can be found. 
We provide a procedure for determining this homogeneous data, thereby closing the entire system of Einstein equations. 
We exemplify this method with two concrete examples, given in the Supplementary material: metric perturbations due to a Hernquist dark matter halo \cite{Cardoso:2021wlq}, and metric perturbations in a linearized Vaidya spacetime due to infalling null dust \cite{Poisson:2009pwt}.
At the end of this Letter, we show how our method connects the NP and the GHZ reconstruction schemes.

Similar to the NP reconstruction hierarchy developed by Ref.~\cite{Li:2026rkf}, the method presented in this paper works for generic {\it extended noncompact sources} and can reconstruct both stationary and dynamical metric perturbations.
This framework will be particularly useful for modeling deformations of black hole spacetimes due to environmental effects, such as dark matter spikes/halo \cite{Cardoso:2021wlq, Hernquist:1990be, Navarro:1995iw} or accretion disks \cite{Abramowicz:2011xu, Shakura:1972te, Novikov:1973kta, Speri:2022upm}, and beyond general-relativity corrections \cite{Will:2014kxa, Lam:2025elw, Lam:2025fzi, Fernandes:2025vxg, Lam:2026grz}, as demonstrated for the first time in Ref.~\cite{Hu:2026ute, Li:2026hpw}.  
Furthermore, this framework can supply crucial input to model dynamical perturbations at nonlinear order or with beyond-vacuum-general-relativity effects, including EMRIs \cite{Amaro-Seoane:2014ela, Barack:2018yly, Upton:2021oxf, Upton:2023tcv, Li:2025ffh, Li:2026foy, Kocsis:2011dr, Yunes:2011ws, Barausse:2007dy, Dyson:2025dlj, Maselli:2020zgv, Maselli:2021men, Yunes:2011aa, Pani:2011xj} and ringdown waveforms \cite{Mitman:2022qdl, Cheung:2023vki, Ma:2024qcv, Khera:2024bjs, Khera:2026lna, Berti:2025hly, Chung:2025gyg, Chung:2024ira, Chung:2024vaf, Laeuger:2025zgb, Li:2022pcy, Hussain:2022ins, Cano:2023tmv, Li:2025fci, Speeney:2024mas, Pezzella:2024tkf}.

 %---------------------------------------------------------------------
\vspace{0.2cm}
\noindent 
\textit{Traceful radiation gauges.}\textemdash \
The solution to the linearized Einstein equations is subject to gauge freedom from diffeomorphisms. 
A convenient choice of gauge is the traceful radiation gauge \cite{Li:2026rkf}, where we fix 
\begin{align}
\label{eq:IRG-ORG}
    h_{\mu\nu}l^{\mu} = 0 \quad\mathrm{or}\quad h_{\mu\nu}n^{\mu} = 0, 
\end{align}
where $l^{\mu}$ and $n^{\mu}$ are tetrad vectors along the ingoing and outgoing principal null directions, respectively, while $m^{\mu}$ and $\bar{m}^{\mu}$ are null tetrads that span the remaining orthogonal directions. Here and henceforth, Greek letters represent spacetime indices. The first and second relations in Eq.~\eqref{eq:IRG-ORG} are the traceful ingoing and outgoing radiation gauges (IRG and ORG), respectively. 
The traceful radiation gauges share many similarities with the Newman-Unti gauge and Bondi-Sachs gauge \cite{Newman:1962cia, Bondi:1962px, Sachs:1962wk, Spiers:2026yqx}.

The metric can always be put into any one of the radiation gauges by solving for the gauge vector $\xi^{\mu}$ \cite{Li:2026rkf, Price:2006ke}
\begin{equation}
\begin{split}
    \left[h_{\mu\nu} - 2\nabla_{(\mu}\xi_{\nu)}\right] l^{\mu} = 0 \ \ \mathrm{or} \ \
    \left[h_{\mu\nu} - 2\nabla_{(\mu}\xi_{\nu)}\right] n^{\mu} = 0.
\end{split}
\end{equation}
In this Letter, we will adopt the ORG, where $h_{ln}$, $h_{nn}$, $h_{nm}$ and $h_{n\bar{m}}$ all vanish; here and henceforth, Latin letters represent tetrad indices. Previously, the traceful ORG was used in \cite{Li:2026rkf} to extend the NP direct reconstruction in \cite{Loutrel:2020wbw, Ripley:2020xby} to generic sources.
Here, we find that the same gauge simplifies the linearized Einstein operator $\cal E$, allowing us to reconstruct the metric by directly inverting it, hierarchically. 
Equivalently, since we assumed the background is of Petrov type-D, one could also impose the IRG; the set of IRG equations and solutions will be the GHP-primed version of what is stated here.
% \KL{Since we can set the leading $r^{-1}$ term in $h_{m\bar{m}}$ to zero by a residual gauge transformation in traceful ORG, the coordinate radius $r$ remains to be the areal radius to leading order in $r$. This enforces the remaining gauge condition for the Bondi gauge! Therefore, the traceful ORG is very much a linearized advanced Bondi gauge adopted to include $T_{ab}$. }

%---------------------------------------------------------------------
\vspace{0.2cm}
\noindent 
\textit{Einstein Reconstruction.}\textemdash \
The reconstruction procedure follows a hierarchical structure, where we first compute $h_{m\bar{m}}$ and $h_{\bar{m}\bar{m}}$, then $h_{l\bar{m}}$, and lastly $h_{ll}$. 
To compute the spin-0 trace component $h_{m\bar{m}}$, we invert the $({\cal E}h)_{nn}$ component of the Einstein equations  [Eq.~\eqref{eq:Enn_full}], which,
in the traceful ORG, leads to
\begin{align}\label{eq:hmmb}
    {\cal E}_{nn}^{(m\bar{m})} h_{m\bar{m}} = -\left[\thorn'(\thorn' - \rho' - \bar{\rho}') + 2\rho'\bar{\rho}'\right] h_{m\bar{m}} = T_{nn}\,,
\end{align}
where ${\cal E}_{ab}^{(cd)}$ is the differential operator acting on $h_{cd}$ in the Einstein equation component $({\cal E}h)_{ab}$, the specific form of which is given in the Supplementary Material [Sec.~\ref{sec:Operator}]. In this equation and henceforth, we use the metric signature $(-,+,+,+)$ and set $8\pi G = 1 = c$.

Given $\Psi_4$ as the solution to the Teukolsky equation, 
\begin{equation}\label{eq:TeukolskyEq}
    {\cal O}_4\Psi_4 = {}_{-2}T, 
\end{equation}
we invert the relation between $\Psi_4$ and $h_{ab}$ given in Eq.~\eqref{eq:T4} to reconstruct the spin-2 component $h_{\bar{m}\bar{m}}$.
In the traceful ORG, $h_{nb} = 0$, so $h_{\bar{m}\bar{m}}$ satisfies a second-order transport equation, 
\begin{align}\label{eq:hmbmb}
    {\cal T}_4^{(\bar{m}\bar{m})} h_{\bar{m}\bar{m}} = -\frac{1}{2} (\thorn'-\bar{\rho}')(\thorn'-\bar{\rho}') h_{\bar{m}\bar{m}} = \Psi_4, 
\end{align}
which can be readily integrated along $n^{\mu}$. 
Next, to find the spin-1 component $h_{l\bar{m}}$, we find that the $(n,\bar{m})$ component of the Einstein equations [Eq.~\eqref{eq:Enmb_full}] acts as a second-order transport equation for $h_{l\bar{m}}$: 
\begin{align}\label{eq:hlmb}
    {\cal E}_{n\bar{m}}^{(l\bar{m})} h_{l\bar{m}} = T_{n\bar{m}} - {\cal E}_{n\bar{m}}^{(\bar{m}\bar{m})} h_{\bar{m}\bar{m}} - {\cal E}_{n\bar{m}}^{(m\bar{m})} h_{m\bar{m}}, 
\end{align}
where the effective source on the right-hand side can be computed using solutions of Eqs.~\eqref{eq:hmmb} and \eqref{eq:hmbmb}.
Lastly, to reconstruct $h_{ll}$, we use the $(l,n)$ component of the Einstein equation [Eq.~\eqref{eq:Eln_full}]\footnote{The $(m,\bar{m})$ component of the Einstein equation [Eq.~\eqref{eq:Emmb_full}] can also be used, which is a second-order transport equation for $h_{ll}$. We found that these two routes are equivalent, as the satisfaction of one implies the other, according to the Bianchi identities.}, which is a first-order transport equation for $h_{ll}$, 
\begin{align}\label{eq:hll} 
    {\cal E}_{ln}^{(ll)} h_{ll} &= T_{ln} - {\cal E}_{ln}^{(lm)} h_{lm} - {\cal E}_{ln}^{(l\bar{m})} h_{l\bar{m}} \nonumber \\
    &\quad - {\cal E}_{ln}^{(mm)} h_{mm} - {\cal E}_{ln}^{(\bar{m}\bar{m})} h_{\bar{m}\bar{m}} - {\cal E}_{ln}^{(m\bar{m})} h_{m\bar{m}}.
\end{align}
Equations~\eqref{eq:hmmb}, \eqref{eq:hmbmb}, \eqref{eq:hlmb} and \eqref{eq:hll} form a complete system that allows us to reconstruct the metric perturbation $h_{ab}$ in the traceful ORG. 
These equations can be readily integrated using Held's integration technique, summarized in the Supplementary Material [Sec.~\ref{sec:Held}], and the complete solution is presented below in Eq.~\eqref{eq:CompleteSolution}. 

%---------------------------------------------------------------------
\vspace{0.2cm}
\noindent 
\textit{Satisfaction of all linearized Einstein equations.}\textemdash \
In the Einstein reconstruction approach, we solve four of the ten Einstein equations, namely $({\cal E}h)_{nn}$, $({\cal E}h)_{n\bar{m}}$, and $({\cal E}h)_{ln}$, together with their complex conjugates. 
It remains to show that the reconstructed metric satisfies the other six components of the Einstein equations.
We find that this follows straightforwardly from Wald's identity \cite{Wald:1978vm} and the Bianchi identities in GHP form, both given in the Supplementary Material [Sec.~\ref{sec:Operator}]. 

\vspace{0.2cm}
\noindent
\textbf{Theorem 1.} \textit{Suppose $h_{ab}$ is the metric perturbation reconstructed from Eqs.~\eqref{eq:hmbmb}--\eqref{eq:hll} on a vacuum Petrov type-D background with $\mathrm{Re}\,\rho' \neq 0$ and the stress-energy tensor $T_{ab}$ is conserved.
If, on some hypersurface $\Sigma$ that is transverse to $n^\mu$, the Held scalars $f_{ll}^{\circ}$, $f_{l\bar{m}}^{\circ}$, $f_{\bar{m}\bar{m}}^{\circ}$ and $g_{\bar{m}\bar{m}}^{\circ}$ (defined below) all vanish, then the full set of Einstein equations are satisfied in $D(\Sigma)$, the region swept out by the integral curves of $n^{\mu}$ that intersect $\Sigma$. } \\

\noindent
\textit{Proof.}---
First, the Wald identity allows us to relate Eq.~\eqref{eq:hmbmb} to the $(\bar{m},\bar{m})$ component of the Einstein equations via
\begin{align}
    {\cal S}_4({\cal E}h) = {\cal O}_4({\cal T}_4h) = {\cal O}_4 \Psi_4 = {}_{-2}T. 
\end{align}
The first equality follows from Wald, the second from Eq.~\eqref{eq:hmbmb}, and the third from Eq.~\eqref{eq:TeukolskyEq}. 
Thus, ${\cal S}_4F = 0$, with $F_{ab} = ({\cal E}h)_{ab} - T_{ab}$.
In traceful ORG gauge, we find 
\begin{align}\label{eq:WaldORG}
    (\thorn' - 4\rho' - \bar{\rho}')(\thorn' - \bar{\rho}')F_{\bar{m}\bar{m}} &= 0, 
\end{align}
and using Held's integration technique, we find 
\begin{equation}\label{eq:Fmbmb_eq}
    F_{\bar{m}\bar{m}} = f_{\bar{m}\bar{m}}^{\circ} \rho'^3\bar{\rho}' + g_{\bar{m}\bar{m}}^{\circ} \bar{\rho}', 
\end{equation}
where $f_{\bar{m}\bar{m}}^{\circ}$ and $g_{\bar{m}\bar{m}}^{\circ}$ are Held scalars that are constant along $n^\mu$. 
If one imposes that, on some hypersurface $\Sigma$ that is transverse to $n^{\mu}$, $f_{\bar{m}\bar{m}}^{\circ}|_{\Sigma} = g_{\bar{m}\bar{m}}^{\circ}|_{\Sigma} = 0$, then the $(\bar{m},\bar{m})$ component of the Einstein equations is satisfied in $D(\Sigma)$.

Next, we use the Bianchi identity to show that the remaining four Einstein equations will also be satisfied under suitable conditions.
First, since $\nabla^b({\cal E}h)_{ab} = 0$ for any symmetric tensor $h_{ab}$, and the stress-energy tensor is conserved $\nabla^b T_{ab} = 0$, then $\nabla^b F_{ab} = 0$ for any $h_{ab}$. 
Projecting this onto $n^a$, Eq.~\eqref{eq:BianchiFull} shows that $(\mathrm{Re}\,\rho') F_{m\bar{m}} = 0$. 
Assuming that $\mathrm{Re}\,\rho' \neq 0$, as $F_{nb} = 0$, 
\begin{equation}\label{eq:Fmmb_eq}
    F_{m\bar{m}} = 0.
\end{equation}
Next, projecting the Bianchi identity onto $\bar{m}^a$ and if $f_{\bar{m}\bar{m}}^{\circ}|_{\Sigma} = g_{\bar{m}\bar{m}}^{\circ}|_{\Sigma} = 0$, we have
\begin{equation}\label{eq:Flmb_eq}
    (\thorn' - 2\rho' - \bar{\rho}')F_{l\bar{m}} = 0 \quad\Rightarrow\quad F_{l\bar{m}} = f_{l\bar{m}}^{\circ} \rho'^2\bar{\rho}'
\end{equation}
in $D(\Sigma)$. 
If $f_{l\bar{m}}^{\circ}|_{\Sigma}$ also vanishes, then the projection onto $l^a$ gives
\begin{equation}\label{eq:Fll_eq}
    (\thorn' - \rho' - \bar{\rho}')F_{ll} = 0 \quad\Rightarrow\quad F_{ll} = f_{ll}^{\circ} \rho' \bar{\rho}'.
\end{equation}
Requiring $f_{ll}^{\circ}|_{\Sigma} = 0$ implies that $F_{ll} = 0$ in $D(\Sigma)$. 
Therefore, if $f_{ab}^{\circ}$ and $g_{\bar{m}\bar{m}}^{\circ}$ all vanish on some hypersurface $\Sigma$, the full set of Einstein equations are satisfied in $D(\Sigma)$. \qed

These operator identities allow us to restrict the solution space for the Einstein equations to just seven functional degrees of freedom, encoded in $f_{ab}^{\circ}$ and $g_{\bar{m}\bar{m}}^{\circ}$. 
A similar analysis using the Bianchi identities was also performed independently in \cite{Li:2026rkf} for the NP reconstruction approach, which showed that their reconstruction approach also closes the full set of Einstein equations when given suitable initial data on $\Sigma$.

%---------------------------------------------------------------------
\vspace{0.2cm}
\noindent 
\textit{Homogeneous data for Kerr.}\textemdash \
Given a solution computed with the Einstein reconstruction scheme, how would one ensure that $f_{ab}^{\circ}|_{\Sigma} = g_{\bar{m}\bar{m}}^{\circ}|_{\Sigma} = 0$ in the first place?
In the Einstein reconstruction procedure, when the transport equations are integrated along $n^{\mu}$, there is additional freedom coming from the homogeneous data. 
It is conceivable that, by choosing the correct homogeneous solution for $h_{ab}$, one can set $f_{ab}^{\circ} = g_{\bar{m}\bar{m}}^{\circ} = 0$.
We have found that there is a systematic procedure to set the homogeneous data of $h_{ab}$ on the Kerr background in ingoing Eddington-Finkelstein coordinates $(v,r,\theta,\varphi)$ with the boosted Kinnersley tetrad \cite{Kinnersley:1969zza}
\begin{align}\label{eq:tetrad}
\begin{split}
    l^{\mu} &= \frac{1}{\Gamma\bar{\Gamma}}\left(r^2 + a^2, \frac{\Delta}{2}, 0, a\right), \\
    n^{\mu} &= (0, -1, 0, 0), \\
    m^{\mu} &= \frac{1}{\sqrt{2}\Gamma}\left(ia\sin\theta, 0, 1, i\csc\theta\right), 
\end{split}
\end{align}
where $\Delta = r^2 - 2Mr + a^2$ and $\Gamma = r + ia\cos\theta$. 
Immediately, one finds that $\rho' = \Gamma^{-1} = O(r^{-1})$. 
If $T_{\bar{m}\bar{m}}$ decays faster than $r^{-1}$ and $h_{ab} = O(r^{-1})$ due to asymptotic flatness, $g_{\bar{m}\bar{m}}^{\circ} = 0$. 
Below, we fix the remaining $f_{ab}^{\circ}$.  

Integrating Eqs.~\eqref{eq:hmmb}, \eqref{eq:hmbmb}, \eqref{eq:hlmb} and \eqref{eq:hll}, the exact solution is 
\begin{widetext}
\begin{equation}\label{eq:CompleteSolution}
\begin{split}
    h_{\bar{m}\bar{m}} &= d_{\bar{m}\bar{m}}^{\circ} \bar{\rho}' + h_{\bar{m}\bar{m}}^p, \qquad h_{m\bar{m}} = h_{m\bar{m}}^p, \\
    h_{l\bar{m}} &= d_{l\bar{m}}^{\circ} \rho'(\rho' + \bar{\rho}') - \frac{1}{2}(\rho' - \bar{\rho}')\rho'\bar{\tau}'^{\circ} d_{\bar{m}\bar{m}}^{\circ} - \frac{\rho'^2 + \bar{\rho}'^2}{4\bar{\rho}'}\tilde{\eth}d_{\bar{m}\bar{m}}^{\circ} + h_{l\bar{m}}^p, \\
    h_{ll} &= d_{ll}^{\circ} (\rho' + \bar{\rho}') + \left[\rho'^2 (2\bar{\rho}'\bar{\tau}'^{\circ}d_{l\bar{m}}^{\circ} + \tilde{\eth}d_{l\bar{m}}^{\circ}) - \frac{\rho'(\rho' + \bar{\rho}')}{4\bar{\rho}'}(2\bar{\rho}'\bar{\tau}'^{\circ}\tilde{\eth}d_{\bar{m}\bar{m}}^{\circ} + \tilde{\eth}\tilde{\eth}d_{\bar{m}\bar{m}}^{\circ}) + \text{c.c.}\right] + h_{ll}^p, 
\end{split}
\end{equation}
\end{widetext}
where $d_{ab}^{\circ}$ are Held scalars. 
Terms containing $d_{ab}^{\circ}$ are the homogeneous solutions of the transport equations, while $h_{ab}^p$ represents the particular solution that is sourced by $\Psi_4$ and $T_{ab}$, given explicitly in the End Matter (Sec.~\ref{sec:ParticularSolutions}). 
Some Held scalars are eliminated by asymptotic flatness requirements and by residual gauge freedom in the traceful ORG.
Notice that the number of Held scalars $\{d_{\bar{m}\bar{m}}^{\circ}$, $d_{l\bar{m}}^{\circ}$, $d_{ll}^{\circ}\}$ (including complex conjugates) in the homogeneous solution matches the number of remaining Einstein equations $\{F_{\bar{m}\bar{m}}, F_{l\bar{m}}, F_{ll}\}$ that we would like to satisfy. 

Substituting Eq.~\eqref{eq:CompleteSolution} into $F_{\bar{m}\bar{m}}$, $F_{l\bar{m}}$ and $F_{ll}$, we find that 
\begin{align}
    \left(\Omega'^{\circ}\frac{1}{2}\tilde{\eth}'\tilde{\eth} + \tau'^{\circ} \tilde{\eth} + \bar{\tau}'^{\circ}\tilde{\eth}' - \Omega'^{\circ}\bar{\rho}^{\circ} - \Psi'^{\circ}\right) d_{\bar{m}\bar{m}}^{\circ}& \nonumber \\
    - 2\tilde{\eth}' d_{l\bar{m}}^{\circ} + p_{\bar{m}\bar{m}}^{\circ} &= 0 \label{eq:pmbmb} \\
    \left(-\frac{5}{2}\bar{\tau}'^{\circ}\tilde{\thorn} + \frac{1}{4}\Omega'^{\circ}\tilde{\thorn}\tilde{\eth} + 3\bar{\rho}^{\circ}\tilde{\eth} + \frac{1}{2}\tilde{\eth}\tilde{\eth}\tilde{\eth}'\right)d_{\bar{m}\bar{m}}^{\circ}& \nonumber \\
    + 3\tilde{\thorn} d_{l\bar{m}}^{\circ} - \tilde{\eth}'d_{ll}^{\circ} + p_{l\bar{m}}^{\circ} &= 0, \label{eq:plmb} \\
    \tilde{\thorn}\left[2d_{ll}^{\circ} - \tilde{\eth}'\tilde{\eth}'d_{mm}^{\circ} - \tilde{\eth}\tilde{\eth}d_{\bar{m}\bar{m}}^{\circ}\right] + p_{ll}^{\circ} &= 0, \label{eq:pll}
\end{align}
where $p_{ab}^{\circ}$ are defined in Eqs.~\eqref{eq:Fmbmb_p}, \eqref{eq:Flmb_p} and \eqref{eq:Fll_p} by the residual of the Einstein equations due to $h_{ab}^p$. 
If the homogeneous solutions are fixed by the above equations, the complete solution satisfies \textit{all} Einstein equations.

\vspace{0.2cm}
\noindent
\textbf{Theorem 2.} \textit{Suppose $h_{ab}$ is a non-static metric perturbation reconstructed from Eqs.~\eqref{eq:hmbmb}--\eqref{eq:hll} on Kerr with the tetrad in Eq.~\eqref{eq:tetrad}, $h_{ab} = O(r^{-1})$, and the stress-energy tensor $T_{ab}$ is conserved with $T_{\bar{m}\bar{m}}$ decaying faster than $r^{-1}$.
If the homogeneous data is fixed by Eqs.~\eqref{eq:pmbmb}--\eqref{eq:pll}, then the full set of Einstein equations is satisfied. } 

\vspace{0.2cm}
\noindent
\textit{Proof.}---The proof is presented in the End Matter [Sec.~\ref{sec:Theorem2Proof}] due to its length, which includes the derivation of Eqs.~\eqref{eq:pmbmb}--\eqref{eq:pll}. \qed
\vspace{0.2cm}

For general non-static perturbations\footnote{For static perturbations, since Eq.~\eqref{eq:pll} becomes degenerate, we require a stronger decay condition on $T_{ab}$, for which we can set all $p_{ab}^{\circ} = 0$. See Sec.~\ref{sec:Theorem2Proof} of End Matter for details.}, the system in Eqs.~\eqref{eq:pmbmb}--\eqref{eq:pll} can be decoupled, which yields
\begin{equation}\label{eq:pmbmb_decoupled}
    \tilde{\thorn}\tilde{\eth}'\tilde{\eth}'\tilde{\eth}'\tilde{\eth}'d_{mm}^{\circ} + 3\Psi'^{\circ}\tilde{\thorn}\tilde{\thorn}d_{\bar{m}\bar{m}}^{\circ} = \tilde{\eth}'\tilde{\eth}'p_{ll}^{\circ} + 2\tilde{\thorn}\tilde{\eth}'p_{l\bar{m}}^{\circ} + 3\tilde{\thorn}\tilde{\thorn}p_{\bar{m}\bar{m}}^{\circ}, 
\end{equation}
and its complex conjugate.
Once we solve this decoupled equation for $d_{\bar{m}\bar{m}}^{\circ}$, the homogeneous data $d_{l\bar{m}}^{\circ}$ and $d_{ll}^{\circ}$ can be computed via Eqs.~\eqref{eq:plmb} and \eqref{eq:pll}.

Equation~\eqref{eq:pmbmb_decoupled}, together with its complex conjugate, can be solved using the Teukolsky-Starobinsky identity via mode decomposition \cite{Hollands:2024iqp}, as the angular tilde operators $\tilde{\eth}$ and $\tilde{\eth}'$ acting on the mode decomposition become the Chandrasekhar operators ${\cal L}_{-s}^{\dagger}$ and ${\cal L}_s$ \cite{Chandrasekhar:1985kt, Pound:2021qin, Dolan:2023enf, Casals:2026man}, respectively, 
\begin{equation}\label{eq:Chandrasekhar}
\begin{split}
    \tilde{\eth} &= \frac{1}{\sqrt{2}}{\cal L}_{-s}^{\dagger} = \frac{1}{\sqrt{2}}\left(\frac{\partial}{\partial\theta} - m\csc\theta + a\omega\sin\theta - s\cot\theta\right), \\
    \tilde{\eth}'&= \frac{1}{\sqrt{2}}{\cal L}_s = \frac{1}{\sqrt{2}}\left(\frac{\partial}{\partial\theta} + m\csc\theta - a\omega\sin\theta + s\cot\theta\right), 
\end{split}
\end{equation}
and $\tilde{\thorn} = -i\omega$ when acting on Held scalars. 
The mode decomposition of Held scalars reads
\begin{equation}
    d_{ab}^{\circ} = \left(\frac{\bar{\rho}'}{\rho'}\right)^s \sum_{\ell \geq |s|} \sum_{m = -\ell}^{\ell} \int d\omega \, e^{-i\omega v} {}_{s}S_{\ell m}(\theta, \varphi; a\omega) \, d_{ab}^{\ell m}(\omega), 
\end{equation}
where ${}_{s}S_{\ell m}(\theta, \varphi)$ are the spin-weighted spheroidal harmonics, and the Teukolsky-Starobinsky identity reads
\begin{equation}
    \tilde{\eth}'\tilde{\eth}'\tilde{\eth}'\tilde{\eth}'{}_{2}S_{\ell m}(\theta,\varphi; a\omega) = \frac{D_{\ell m}(a\omega)}{4} {}_{-2}S_{\ell m}(\theta,\varphi; a\omega), 
\end{equation}
where $[D_{\ell m}(a\omega)]^2 = \lambda^2(\lambda + 2)^2 + 8a\omega(m - a\omega)\lambda(5\lambda + 6) + 48a^2\omega^2(2\lambda + 3(m - a\omega)^2)$ is the angular Teukolsky-Starobinsky constant \cite{Teukolsky:1974yv, Starobinskii:1974nkd, Berens:2024czo} and $\lambda$ is the eigenvalue of the spin $-2$ spheroidal equation. 
Since $\tilde{\thorn} = -i\omega$, Eq.~\eqref{eq:pmbmb_decoupled} and its complex conjugate reduce, for each $(\ell, m, \omega)$, to a $2\times 2$ algebraic system for $d_{\bar{m}\bar{m}}^{\ell m}(\omega)$ and $d_{mm}^{\ell m}(\omega)$ with determinant $\omega^2\left([D_{\ell m}(a\omega)]^2 + 144\Psi'^{\circ\,2}\omega^2\right)/16 = \omega^2|C_{\ell m}|^2/16$, where $|C_{\ell m}|^2$ is the Teukolsky-Starobinsky constant \cite{Teukolsky:1974yv, Starobinskii:1974nkd}, so Eqs.~\eqref{eq:pmbmb}--\eqref{eq:pll} are uniquely solvable for all $\omega \neq 0$ away from the algebraically special frequencies at which $|C_{\ell m}| = 0$ \cite{Chandrasekhar:1985kt}. 

For the residual terms $p_{ab}^{\circ}$, since the equations for homogeneous data are independent of which hypersurface $\Sigma$ we evaluate on, one could compute them on any hypersurface that we find convenient, e.g., at past infinity $\mathscr{I}^-$. 
Then, the asymptotic properties of $T_{ab}$ and $\Psi_4$ fully determine the homogeneous solution. 
The asymptotic expansions of Eqs.~\eqref{eq:pmbmb}--\eqref{eq:pll} match those in \cite{Hollands:2024iqp} after converting to the ORG, which are related to Bondi flux balance laws in the ORG \cite{Wald:1984rg}. 

%---------------------------------------------------------------------
\vspace{0.2cm}
\noindent 
\textit{Connection to other metric reconstruction schemes.}\textemdash \
Our method is closely related to the recently developed NP reconstruction method for generic sources \cite{Li:2026rkf}, and can be thought of as another route within the NP reconstruction scheme. 
In fact, the first two steps in our scheme are identical to theirs when written in GHP form.
When computing $h_{\bar{m}\bar{m}}$, the first-order transport equation given in \cite{Li:2026rkf} is a factorized form of the second-order transport equation [Eq.~\eqref{eq:hmbmb}].
Similarly, the trace component $h_{m\bar{m}}$ also has the same structure. 
However, when computing $h_{l\bar{m}}$, \cite{Li:2026rkf} used one of the Bianchi identities to first compute $\Psi_3$, then $\tau'^{(1)}$ using Eq.~\eqref{eq:ThornTau}, and lastly $h_{l\bar{m}}$ using Eq.~\eqref{eq:tau1}. 
In our scheme, we have used $({\cal E}h)_{n\bar{m}}$ instead, which was derived by summing Eqs.~\eqref{eq:ThornTau} and \eqref{eq:EthRho}, and expressing the linearized spin coefficients in $h_{ab}$ using Eqs.~\eqref{eq:rho1} and \eqref{eq:tau1}. 
Finally, for the $h_{ll}$ component, \cite{Li:2026rkf} first computed $\Psi_2$ using one of the Bianchi identities, and used Eq.~\eqref{eq:ThornRhoEthTau} to find $h_{ll}$. 
In our case, the final step involves $({\cal E}h)_{ln}$, which can be derived from combining Eqs.~\eqref{eq:ThornThornCommutator} and \eqref{eq:ThornRhoEthTau}.

Since the NP equations are integrable \cite{Papapetrou:1971b, Edgar:1980tso}, the metric reconstructed from these two procedures will yield the same solution.
However, with our scheme, since we did not use the GHP Bianchi identities throughout the reconstruction process, these identities can be used separately to verify whether the solution satisfies all 10 Einstein equations. 
This is advantageous because the Bianchi identities restrict the $\rho'$-dependence of the residual of the Einstein equations, and thus, they allow us to fix the homogeneous data. 
More generally, the Einstein reconstruction scheme provides a second-order reconstruction procedure, while the NP reconstruction scheme is essentially a first-order formulation of the former. 
The former is more direct for metric reconstruction since it does not require intermediate NP quantities, such as the perturbed spin coefficients and Weyl scalars, which depend on how the tetrad perturbations are defined [Eq.~\eqref{eq:tetrad_pert}]. 
The latter approach may be advantageous if NP quantities are also needed for subsequent calculations, such as in the modified Teukolsky formalism \cite{Li:2022pcy, Li:2025fci, Cano:2023tmv}.

Another metric reconstruction scheme for nonvacuum perturbations is the GHZ scheme \cite{Green:2019nam}, as discussed in the Introduction.
In this approach, the Hertz-potential reconstruction is supplemented by a corrector tensor
$x_{ab} = 2m_{(a}\bar{m}_{b)} x_{m\bar{m}} - 2n_{(a}\bar{m}_{b)} x_{lm} - 2n_{(a}m_{b)} x_{l\bar{m}} + n_{a}n_{b} x_{ll}$ that satisfies 
\begin{equation}\label{eq:CorrectorTensor}
    [({\cal E}x)_{ab} - T_{ab}]n^b = 0, 
\end{equation}
such that the metric perturbation sourced by $T_{ab}$ can be computed. 
We have here converted the original formulation in \cite{Green:2019nam} into the ORG for direct comparison with our work. 
In their scheme, any metric perturbation satisfying the linearized Einstein equations with compact sources is shown to admit the following decomposition:
\begin{equation}
    h_{ab} = \mathfrak{h}_{ab} + x_{ab} + \dot{g}_{ab} + ({\cal L}_{\xi} g)_{ab}, 
\end{equation}
where $\mathfrak{h}_{ab} = \mathrm{Re}({\cal S}_4^{\dagger}\Phi)$ is the Hertz-reconstructed metric perturbation, with $\Phi$ the Hertz potential, $\dot{g}_{ab}$ the zero modes associated with mass and angular momentum shifts, and $({\cal L}_{\xi} g)_{ab}$ the gauge modes. 

To identify how the GHZ scheme is related to the Einstein reconstruction scheme, let us take their decomposition and pass it through our reconstruction procedure. 
First, Eq.~\eqref{eq:hmbmb} yields
\begin{equation}\label{eq:GHZ_hmbmb}
    \Psi_4 = {\cal T}_4[\mathfrak{h}_{\bar{m}\bar{m}}]. 
\end{equation}
The corrector tensor does not enter this equation since $x_{\bar{m}\bar{m}} = 0$, the zero mode does not have a $(\bar{m},\bar{m})$ component \cite{Toomani:2021jlo, Hollands:2024iqp}, and gauge modes do not affect the gauge-invariant $\Psi_4$. 
The Einstein components [Eqs.~\eqref{eq:hmmb}--\eqref{eq:hll}] give us
\begin{equation}
    ({\cal E} \mathfrak{h})_{nb} + ({\cal E} x)_{nb} = T_{nb}, 
\end{equation}
Note that the zero modes and gauge modes are annihilated by $\cal E$. 
By construction, the corrector tensor is obtained by solving the $(n,b)$ components of the Einstein equations. 
Hence, $({\cal E} x)_{nb}$ cancels out $T_{nb}$ exactly and we have 
\begin{equation}\label{eq:hbar_nmu}
    ({\cal E}\mathfrak{h})_{nb} = 0.
\end{equation}
The solution to Eqs.~\eqref{eq:GHZ_hmbmb} and \eqref{eq:hbar_nmu} is just the metric perturbation given in Eq.~\eqref{eq:CompleteSolution} (modulo zero modes and gauge modes), except that all $T_{ab}$-dependence in the particular solution is removed.

For the other components of the Einstein equations, 
\begin{align}\label{eq:hbar_S}
    ({\cal E}\mathfrak{h})_{ab} = S_{ab}, 
\end{align}
where $S_{ab} = T_{ab} - ({\cal E}x)_{ab}$ and $ab = \{ll, l\bar{m}, \bar{m}\bar{m}\}$. 
These equations, although they seemingly depend on the source through $S_{ab}$, can be satisfied thanks to the Wald identity and Bianchi identity when subject to suitable asymptotic decay and with suitable homogeneous solutions. 
This can be proven via a similar procedure to that described around Eqs.~\eqref{eq:Fmbmb_eq}--\eqref{eq:Fll_eq}.
Therefore, the GHZ decomposition cleanly divides the $T_{ab}$-driven solution and the radiation-driven solution.

On the other hand, if we continue with the GHZ decomposition and suppose there exist $\Phi$ and $\eta$ such that
\begin{gather}
    {\cal O}_4^{\dagger} \Phi = \eta, \quad {\cal T}_4[\mathrm{Re}({\cal S}_4^{\dagger}\Phi)] = \Psi_4, \label{eq:Phi} \\
    \mathrm{Re}({\cal T}_4^{\dagger} \eta)_{ab} = S_{ab}, \label{eq:eta}
\end{gather}
then, the Wald identity implies
\begin{align}
    ({\cal E}\mathfrak{h})_{ab} &= {\cal E}[\mathrm{Re}({\cal S}_4^{\dagger}\Phi)]_{ab} = \mathrm{Re}({\cal E}{\cal S}_4^{\dagger}\Phi)_{ab} \nonumber \\
    &= \mathrm{Re}({\cal T}_4^{\dagger}{\cal O}_4^{\dagger}\Phi)_{ab} = \mathrm{Re}({\cal T}_4^{\dagger}\eta)_{ab} = S_{ab}, 
\end{align}
The solutions of the linearized Einstein equations are unique up to gauge modes \cite{Choquet-Bruhat:2009xil, Fewster:2012bj}. Since we supply the same source $T_{ab}$, the same radiation information from $\Psi_4$, and the same initial/boundary conditions, the GHZ reconstructed solution must be the same as that computed with our Einstein reconstruction scheme, after including the zero modes externally by fixing the ADM charges. 

A potential obstruction for the GHZ decomposition as a metric reconstruction scheme is how to compute $\eta$ and $\Phi$ from the source. 
References \cite{Hollands:2024iqp, Casals:2026man} showed that the GHZ scheme can be formulated as an initial-value problem. If, on a hypersurface lying in a region where $S_{ab} = 0$, one can set $\eta = 0$, then $\Phi$ can be obtained from the Teukolsky-Starobinsky identities \cite{Ori:2002uv}. The quantities
$\eta$ and $\Phi$ can thus be integrated into regions with nonzero $S_{ab}$ using trivial initial data \cite{Green:2019nam, Hollands:2024iqp, Casals:2026man}.
However, for generic sources, this argument is not guaranteed to hold since trivial initial data could be incompatible with the source\footnote{We find that the initial/boundary data for $\eta$ can be found using an off-shell identity: $\nabla^b(\mathrm{Re}{\cal T}_4^{\dagger}\tilde{\eta})_{ab} = 0$ for any $\tilde{\eta}$ with GHP-weight $\{4,0\}$, and using the conservation of $S_{ab}$. 
If we define $\eta$ as the solution of $(\mathrm{Re}{\cal T}_4^{\dagger}\eta)_{\bar{m}\bar{m}} = S_{\bar{m}\bar{m}}$, following a similar approach as in Eqs.~\eqref{eq:Flmb_eq}--\eqref{eq:Fll_eq} and the homogeneous fixing procedure, we can then show that $\eta$ satisfies Eq.~\eqref{eq:eta} by choosing suitable initial/boundary data for $\eta$ in terms of $S_{ab}$. 
However, we were not able to fix the initial/boundary data for $\Phi$, which we leave as future work. }. 
Nevertheless, since the GHZ framework holds, compatible $\eta$ and $\Phi$ must exist. 
Once suitable initial data are given, both methods can be used to compute $h_{ab}$ \cite{Hollands:2024iqp}. 

The Einstein reconstruction scheme and the GHZ approach are complementary. 
Analogous to electrostatics, there are two approaches to finding the electric field. 
The field formulation computes the electric field directly from Gauss' law, while the potential formulation relies on an auxiliary scalar field that satisfies the Poisson equation, whose solution is simple. 
The potential formulation is equivalent to the field formulation for two reasons: (i) the potential, as a solution to the Poisson equation subject to general boundary conditions and charge density (with suitable decay properties), always exists \cite{Jackson:1998nia}; (ii) it is compatible with the condition where the curl of the electric field is zero.
In our case, Einstein reconstruction amounts to the field formulation, where we directly compute $h_{ab}$, while the Hertz-potential-based approaches are analogous to the potential formulation, which is based on the auxiliary Hertz potential $\Phi$. 
This potential $\Phi$ satisfies the sourced adjoint Teukolsky equation, for which solutions generally exist. 
However, in the presence of sources, it remains to be shown how to generically compute a $\Phi$ that is compatible with $\Psi_4$ and $T_{ab}$.

%---------------------------------------------------------------------
\vspace{0.2cm}
\noindent 
\textit{Conclusion and Future Directions.}\textemdash \
The Einstein reconstruction scheme provides a complete method for reconstructing the metric perturbation from $\Psi_4$ and $T_{ab}$. 
The metric solutions driven by $\Psi_4$ and $T_{ab}$ are provided explicitly as integrals along the principal null direction.
Suitable homogeneous data can be fixed systematically using the residual of the Einstein equations due to the particular solutions. 
This completes previous direct reconstruction approaches by showing how the full set of Einstein equations is satisfied and how the homogeneous data is fixed. 
Two concrete examples are provided in the Supplementary material to illustrate how the procedure can be applied to reconstruct static and dynamical perturbations. 
This reconstruction method is a hybrid approach of both the NP and GHZ approaches, thus bridging the two methods. 

The Einstein reconstruction scheme, along with the method proposed in \cite{Li:2026rkf}, can be applied to compute perturbations of noncompact sources, which are ubiquitous in astrophysical environments. 
For example, metric perturbations due to beyond general-relativity corrections \cite{Hu:2026ute, Li:2026hpw} or environmental effects, such as dark matter spikes/halo and accretion disks around black holes. 
The former allows us to accurately compute observables for theory-specific tests of general relativity, such as quasinormal frequencies \cite{Chung:2025gyg, Chung:2024ira, Chung:2024vaf, Li:2025fci, Speeney:2024mas}, while the latter enables accurate waveform modeling, especially for EMRIs where environmental effects were shown to be important \cite{Barausse:2014tra, Dyson:2025dlj, Li:2025ffh, Li:2026foy, Maselli:2020zgv, Maselli:2021men, Speri:2022upm, Speri:2024qak, Barsanti:2026ulr}.
Also, given a way to fix the homogeneous solutions, our method could potentially help reconstruct metric perturbations for EMRIs without temporal tapering of the source \cite{Toomani:2021jlo, Casals:2024ynr}.
This procedure can be extended to higher-order perturbations, whose stress-energy tensors -- composed of couplings of lower-order perturbations -- also have extended support. 
This could help explore nonlinear gravitational turbulence in Kerr spacetime \cite{Redondo-Yuste:2023seq, Ma:2025rnv, Yang:2014tla, Lehner:2026tfe}.

Future work could focus on investigating if other gauges also admit such an iterative reconstruction scheme. 
In traceful radiation gauges, many metric components are eliminated so that transport equations are always integrated along the principal null directions.
By contrast, in other gauges, such as in the Lorenz gauge \cite{Dolan:2021ijg, Dolan:2023enf}, GHP derivatives along other directions would remove the hierarchical structure in the Einstein operator, as also occurs in NP-based approaches \cite{Loutrel:2020wbw, Ripley:2020xby, Suvorov:2019qow, Li:2026rkf}. 
Perhaps, combinations of the Einstein equations (or their derivatives) could simplify the resulting set of equations, yielding decoupled equations for individual metric components that can be solved readily. 
This would prevent ``string singularities'' in the reconstructed metric for distributional sources \cite{Ori:2002uv, Dolan:2023enf, Wardell:2024yoi}. 

Other future work could investigate the relation between the Einstein reconstruction, NP and GHZ methods to another metric reconstruction approach developed by Aksteiner, Andersson and B\"ackdahl~\cite{Aksteiner:2016pjt, Dolan:2023enf, Wardell:2024yoi}.
The latter uses both radiative Weyl scalars to reconstruct the time-derivative of the metric by acting with a second-order differential operator on Weyl scalars. 
Since this method reconstructs only the time-derivative of the metric, static and zero modes cannot be captured by this procedure. 
Nevertheless, in the radiative sector, establishing the connection between this method and both the NP and GHZ approaches could reveal further underlying structure of metric reconstruction.

%---------------------------------------------------------------------
\vspace{0.2cm}
\noindent 
\textit{Acknowledgments.}\textemdash \
The authors thank Roman Berens for insightful discussions on the manuscript. 
The authors acknowledge the support from the Simons Foundation through Award No. 896696, the Simons Foundation International under grant SFI-MPS-BH-00012593-01, the NSF through Grant No. PHY-2207650 and PHY 25-12423. 
Many of the calculations here are done with \texttt{xAct} \cite{MartinGarcia:2002xact, GomezLobo:2009ghp} with \texttt{Mathematica} notebooks developed by \cite{Spiers:2023cip, Spiers:2024src, Spiers2023:ghp}.

\clearpage
%---------------------------------------------------------------------
\onecolumngrid

\section*{End Matter}
\section{Particular Solutions}\label{sec:ParticularSolutions}

The particular solution for Eqs.~\eqref{eq:hmbmb}--\eqref{eq:hll} contained in Eq.~\eqref{eq:CompleteSolution} can be expressed explicitly as integrals along $n^{\mu}$, 
\begin{align}
    h_{\bar{m}\bar{m}}^p &= 2\left[\int d\lambda \frac{\Psi_4}{\bar{\rho}'} - \bar{\rho}'\int d\lambda \frac{\Psi_4}{\bar{\rho}'^2}\right], \label{eq:hmbmb_particular} \\
    h_{m\bar{m}}^p &= \frac{1}{2}\bigg[\left(\frac{\bar{\rho}'}{\rho'} + \frac{\rho'}{\bar{\rho}'}\right)\int d\lambda\,\frac{T_{nn}}{2}\left(\frac{1}{\rho'} + \frac{1}{\bar{\rho}'}\right) - (\rho' + \bar{\rho}')\int d\lambda\,\frac{T_{nn}}{2}\left(\frac{1}{\rho'^2} + \frac{1}{\bar{\rho}'^2}\right)\bigg], \label{eq:hmmb_particular} \\
    h_{l\bar{m}}^p &= -\frac{1}{3}\bigg[\frac{\rho'(\rho' - 2\bar{\rho}')}{\bar{\rho}'^3}\int d\lambda\,\tilde{T}_{n\bar{m}}\frac{\rho' + \bar{\rho}'}{\rho'} - \rho'(\rho' + \bar{\rho}')\int d\lambda\,\tilde{T}_{n\bar{m}}\frac{\rho' - 2\bar{\rho}'}{\rho'\bar{\rho}'^3}\bigg], \label{eq:hlmb_particular} \\
    h_{ll}^p &= -2(\rho' + \bar{\rho}')\int d\lambda\,\frac{\tilde{T}_{ln}}{(\rho' + \bar{\rho}')^2}, \label{eq:hll_particular}
\end{align}
where the effective source terms in Eqs.~\eqref{eq:hlmb_particular} and \eqref{eq:hll_particular} are
\begin{align}
    \tilde{T}_{n\bar{m}} &= T_{n\bar{m}} - {\cal E}_{n\bar{m}}^{(\bar{m}\bar{m})} h_{\bar{m}\bar{m}}^p - {\cal E}_{n\bar{m}}^{(m\bar{m})} h_{m\bar{m}}^p, \\
    \tilde{T}_{ln} &= T_{ln} - {\cal E}_{ln}^{(lm)} h_{lm}^p - {\cal E}_{ln}^{(l\bar{m})} h_{l\bar{m}}^p - {\cal E}_{ln}^{(mm)} h_{mm}^p - {\cal E}_{ln}^{(\bar{m}\bar{m})} h_{\bar{m}\bar{m}}^p - {\cal E}_{ln}^{(m\bar{m})} h_{m\bar{m}}^p.
\end{align}

\section{Proof of Theorem 2}\label{sec:Theorem2Proof}
\subsection{Dynamical Perturbations}

Here, we derive the homogeneous fixing equations that ensure that the remaining Einstein equations are satisfied. 
Integrating Eqs.~\eqref{eq:hmbmb}--\eqref{eq:hll} successively, we find that
\begin{equation}
\begin{split}
    h_{\bar{m}\bar{m}} &= d_{\bar{m}\bar{m}}^{\circ} \bar{\rho}' + h_{\bar{m}\bar{m}}^p, \qquad h_{m\bar{m}} = d_{m\bar{m}}^{\circ}(\rho' + \bar{\rho}') + h_{m\bar{m}}^p, \\
    h_{l\bar{m}} &= d_{l\bar{m}}^{\circ} \rho'(\rho' + \bar{\rho}') - \frac{1}{2}(\rho' - \bar{\rho}')\rho'\bar{\tau}'^{\circ} d_{\bar{m}\bar{m}}^{\circ} - \frac{\rho'^2 + \bar{\rho}'^2}{4\bar{\rho}'}\tilde{\eth}d_{\bar{m}\bar{m}}^{\circ} + \bar{\rho}'\tilde{\eth}'d_{m\bar{m}}^{\circ} + \bar{\rho}'(\rho' + \bar{\rho}')\tau'^{\circ}d_{m\bar{m}}^{\circ} + h_{l\bar{m}}^p, \\
    h_{ll} &= d_{ll}^{\circ} (\rho' + \bar{\rho}') + \left[\rho'^2 (2\bar{\rho}'\bar{\tau}'^{\circ}d_{l\bar{m}}^{\circ} + \tilde{\eth}d_{l\bar{m}}^{\circ}) - \frac{\rho'(\rho' + \bar{\rho}')}{4\bar{\rho}'}(2\bar{\rho}'\bar{\tau}'^{\circ}\tilde{\eth}d_{\bar{m}\bar{m}}^{\circ} + \tilde{\eth}\tilde{\eth}d_{\bar{m}\bar{m}}^{\circ}) + \text{c.c.}\right] \\
    &\quad + 2\tilde{\thorn}d_{m\bar{m}}^{\circ} + 2\rho'\bar{\rho}'(\tau'^{\circ}\tilde{\eth} + \bar{\tau}'^{\circ}\tilde{\eth}')d_{m\bar{m}}^{\circ} + \left[2\rho'\bar{\rho}'(\rho' + \bar{\rho}')\tau'^{\circ}\bar{\tau}'^{\circ} + (\rho'^2\Psi'^{\circ} + \bar{\rho}'^2\bar{\Psi}'^{\circ})\right] d_{m\bar{m}}^{\circ} + h_{ll}^p, 
\end{split}
\end{equation}
along with their complex conjugates, is the complete solution. 
This can be checked against the homogeneous solution quoted in \cite{Casals:2024ynr}, except that we also have terms driven by $h_{\bar{m}\bar{m}}$ and we have dropped terms that are not $O(r^{-1})$. 
In the main text [Eq.~\eqref{eq:CompleteSolution}], we have also set $d_{m\bar{m}}^{\circ} = 0$ as it can be removed by using the residual gauge freedom of the traceful ORG, where $h_{ab} \to h_{ab} + 2 \nabla_{(a}\xi_{b)}$ with $\xi_a = d_{m\bar{m}}^{\circ} n_a$ \cite{Hollands:2024iqp, Price:2006ke}. Notice that the homogeneous solution driven by $d_{ll}^{\circ}$ contains the zero mode for mass shift, and the part driven by $d_{lm}^{\circ}$ and $d_{l\bar{m}}^{\circ}$ also includes the zero mode for angular momentum shift \cite{Toomani:2021jlo, Hollands:2024iqp}. 
The zero modes have $d_{ll}^{\circ} = -\Psi'^{\circ} \delta M/M$ and $d_{l\bar{m}}^{\circ} = -(\tau'^{\circ}\Psi'^{\circ}/2\rho^{\circ})(\delta a/a)$. 

Now, we wish to compute the residual of the Einstein equations for this solution. 
We split this task into two parts: the residual sourced by the homogeneous solution and that sourced by the particular solution $h_{ab}^p$.
Let us first compute the latter. 
Since $h_{ab}^p$ is a solution to the Einstein reconstruction procedure, the Wald identity [Eq.~\eqref{eq:WaldORG}] guarantees that 
\begin{equation}\label{eq:Fmbmb_p}
    F_{\bar{m}\bar{m}}^p = p_{\bar{m}\bar{m}}^{\circ} \rho'^3\bar{\rho}', 
\end{equation}
for some Held scalar $p_{\bar{m}\bar{m}}^{\circ}$. 
Next, the $\bar{m}$-component of the Bianchi identity read
\begin{equation}
    (\thorn' - 2\rho' - \bar{\rho}')F_{l\bar{m}}^p  - (\eth - \tau - \bar{\tau}')F_{\bar{m}\bar{m}}^p = 0.
\end{equation}
The $F_{nb}$ and $F_{m\bar{m}}$ terms vanish because $h_{ab}^p$ satisfies Eqs.~\eqref{eq:hmbmb}--\eqref{eq:hll} and \eqref{eq:Fmmb_eq}. 
Substituting in Eq.~\eqref{eq:Fmbmb_p} and integrating along $n^{\mu}$, we find 
\begin{equation}\label{eq:Flmb_p}
    F_{l\bar{m}}^p = p_{l\bar{m}}^{\circ} \rho'^2\bar{\rho}' + \rho'^2\bar{\rho}'\left(\rho'\tilde{\eth}p_{\bar{m}\bar{m}}^{\circ} - 2\rho'^2\bar{\tau}'^{\circ}p_{\bar{m}\bar{m}}^{\circ}\right), 
\end{equation}
for some Held scalar $p_{l\bar{m}}^{\circ}$. 
Lastly, the $l$-component of the Bianchi identity reads
\begin{equation}
    (\thorn' - \rho' - \bar{\rho}')F_{ll}^p - (\eth' - \tau' - 2\bar{\tau})F_{lm}^p - (\eth - 2\tau - \bar{\tau}')F_{l\bar{m}}^p = 0, 
\end{equation}
and 
\begin{equation}\label{eq:Fll_p}
    F_{ll}^p = p_{ll}^{\circ} \rho'\bar{\rho}' + \rho'\bar{\rho}'\left(\rho'\tilde{\eth}p_{l\bar{m}}^{\circ} + \rho'^4\bar{\tau}'^{\circ 2}p_{\bar{m}\bar{m}}^{\circ} - \rho'^3\bar{\tau}'^{\circ}\tilde{\eth}p_{\bar{m}\bar{m}}^{\circ} + \frac{1}{2}\rho'^2\tilde{\eth}\tilde{\eth}p_{\bar{m}\bar{m}}^{\circ} + \mathrm{c.c.}\right),
\end{equation}
for some Held scalar $p_{ll}^{\circ}$. 
Next, to compute $F_{ab}^h$, we directly substitute the homogeneous solution into the Einstein operators. For example, 
\begin{align}
    F_{\bar{m}\bar{m}}^h &= \rho'^3\bar{\rho}' \left[-2\tilde{\eth}' d_{l\bar{m}}^{\circ} - \left(-\Omega'^{\circ}\frac{1}{2}\tilde{\eth}'\tilde{\eth} - \tau'^{\circ} \tilde{\eth} - \bar{\tau}'^{\circ}\tilde{\eth}' + \Omega'^{\circ}\bar{\rho}^{\circ} + \Psi'^{\circ}\right) d_{\bar{m}\bar{m}}^{\circ} \right] 
\end{align}
We omit $F_{l\bar{m}}^h$ and $F_{ll}^h$ since they are extraordinarily lengthy.
Setting $F_{\bar{m}\bar{m}}^h + F_{\bar{m}\bar{m}}^p = 0$, we have 
\begin{equation}\label{eq:pmbmb_appendix}
    -2\tilde{\eth}' d_{l\bar{m}}^{\circ} - \left(-\Omega'^{\circ}\frac{1}{2}\tilde{\eth}'\tilde{\eth} - \tau'^{\circ} \tilde{\eth} - \bar{\tau}'^{\circ}\tilde{\eth}' + \Omega'^{\circ}\bar{\rho}^{\circ} + \Psi'^{\circ}\right) d_{\bar{m}\bar{m}}^{\circ} + p_{\bar{m}\bar{m}}^{\circ} = 0. 
\end{equation}
Next, $F_{l\bar{m}}^h + F_{l\bar{m}}^p = 0$ implies that
\begin{equation}\label{eq:plmb_appendix}
    3\tilde{\thorn} d_{l\bar{m}}^{\circ} - \left(\frac{5}{2}\bar{\tau}'^{\circ}\tilde{\thorn} - \frac{1}{4}\Omega'^{\circ}\tilde{\thorn}\tilde{\eth} - 3\bar{\rho}^{\circ}\tilde{\eth} - \frac{1}{2}\tilde{\eth}\tilde{\eth}\tilde{\eth}'\right)d_{\bar{m}\bar{m}}^{\circ} - \tilde{\eth}'d_{ll}^{\circ} + p_{l\bar{m}}^{\circ} = 0, 
\end{equation}
where $p_{\bar{m}\bar{m}}$ in Eq.~\eqref{eq:Flmb_p} is eliminated using Eq.~\eqref{eq:pmbmb_appendix}.
Lastly, $F_{ll}^h + F_{ll}^p = 0$ implies that
\begin{equation}\label{eq:pll_appendix}
    \tilde{\thorn}\left[2d_{ll}^{\circ} - \tilde{\eth}'\tilde{\eth}'d_{mm}^{\circ} - \tilde{\eth}\tilde{\eth}d_{\bar{m}\bar{m}}^{\circ}\right] + p_{ll}^{\circ} = 0, 
\end{equation}
where $p_{\bar{m}\bar{m}}^{\circ}$ and $p_{l\bar{m}}^{\circ}$ in Eq.~\eqref{eq:Fll_p} are eliminated using Eqs.~\eqref{eq:pmbmb_appendix} and \eqref{eq:plmb_appendix}.
This concludes the derivation of Eq.~\eqref{eq:pmbmb}--\eqref{eq:pll}. \qed 

To decouple the equations, we can take suitable combinations of them. 
Taking $\tilde{\eth}'$ of Eq.~\eqref{eq:plmb_appendix} and subsequently eliminating $d_{l\bar{m}}^{\circ}$ using Eq.~\eqref{eq:pmbmb_appendix}, we have 
\begin{equation}
    \tilde{\eth}'\tilde{\eth}'d_{ll}^{\circ} = -\frac{3}{2}\Psi'^{\circ}\tilde{\thorn}d_{\bar{m}\bar{m}}^{\circ} + \frac{1}{2}\tilde{\eth}'\tilde{\eth}'\tilde{\eth}\tilde{\eth}d_{\bar{m}\bar{m}}^{\circ} + \frac{3}{2}\tilde{\thorn}p_{\bar{m}\bar{m}}^{\circ} + \tilde{\eth}'p_{l\bar{m}}^{\circ}. 
\end{equation}
Taking $\tilde{\eth}'\tilde{\eth}'$ of Eq.~\eqref{eq:pll_appendix} and eliminating $d_{ll}^{\circ}$ using the above equation, we have 
\begin{equation}
    \tilde{\thorn}\tilde{\eth}'\tilde{\eth}'\tilde{\eth}'\tilde{\eth}'d_{mm}^{\circ} + 3\Psi'^{\circ}\tilde{\thorn}\tilde{\thorn}d_{\bar{m}\bar{m}}^{\circ} = \tilde{\eth}'\tilde{\eth}'p_{ll}^{\circ} + 2\tilde{\thorn}\tilde{\eth}'p_{l\bar{m}}^{\circ} + 3\tilde{\thorn}\tilde{\thorn}p_{\bar{m}\bar{m}}^{\circ}.
\end{equation}
This equation can then be solved using the Teukolsky-Starobinsky identity with mode decomposition. 
However, note that applying $\tilde{\eth}'$ to Eqs.~\eqref{eq:plmb_appendix} and \eqref{eq:pll_appendix} discards the components of these equations in the kernel of $\tilde{\eth}'$.
Explicitly, in a Schwarzschild background, they are the $\ell = 1$ and $\ell = 0$ spin-weighted spherical harmonics. 
After solving the decoupled equation for $d_{\bar{m}\bar{m}}^{\circ}$, these components can be recovered by solving Eqs.~\eqref{eq:plmb_appendix} and \eqref{eq:pll_appendix} directly.

\subsection{Static Perturbations}

For static perturbations, $T_{ll}$ cannot decay as $r^{-2}$, as it introduces a mass flux to the system that causes a time-varying ADM mass. 
Therefore, Eq.~\eqref{eq:Fll_p} implies $p_{ll}^{\circ} = 0$, and Eq.~\eqref{eq:pll_appendix} becomes
\begin{equation}
    \tilde{\thorn}\left[-2d_{ll}^{\circ} + \tilde{\eth}'\tilde{\eth}'d_{mm}^{\circ} + \tilde{\eth}\tilde{\eth}d_{\bar{m}\bar{m}}^{\circ}\right] = 0. 
\end{equation}
This equation is automatically true since $d_{ab}^{\circ}$ is static, thus acting $\tilde{\thorn}$ on it yields zero. 
This means we lose one of the equations for determining $d_{ab}^{\circ}$, and the system no longer closes itself. 

To bypass solving the homogeneous fixing equations, we impose a stronger decay property on the source: 
\begin{equation}
    T_{ll} = O(r^{-2-\epsilon}), \qquad T_{ln}, T_{l\bar{m}} = O(r^{-3-\epsilon}), \qquad T_{nn}, T_{n\bar{m}}, T_{\bar{m}\bar{m}} = O(r^{-4-\epsilon}), 
\end{equation}
where $0 < \epsilon < 1$, as a sufficient yet not necessary condition. 
Since the homogeneous solution of $\Psi_4$ does not give a horizon-regular and asymptotically flat $h_{ab}$, only the particular solution remains, and $\Psi_4 = O(r^{-4-\epsilon})$.
The particular solutions $h_{ab}^p$ computed by integrating from $\mathscr{I}^{-}$ has the following decay properties:
\begin{equation}
    h_{ll}^p, h_{m\bar{m}}^p = O(r^{-1-\epsilon}), \qquad h_{\bar{m}\bar{m}}^p, h_{l\bar{m}}^p = O(r^{-2-\epsilon}).
\end{equation}
This implies that 
\begin{equation}
    F_{ll}^p = O(r^{-2-\epsilon}), \qquad F_{l\bar{m}}^p = O(r^{-3-\epsilon}), \qquad F_{\bar{m}\bar{m}}^p = O(r^{-4-\epsilon}), 
\end{equation}
and thus all $p_{ab}^{\circ} = 0$. 
Therefore, the particular solution already satisfies all linearized Einstein equations, and we can simply set $d_{ab}^{\circ} = 0$. 
The remaining degrees of freedom come from zero and gauge modes, as discussed at the start of this Section. 
This was explicitly done in \cite{Hu:2026ute, Li:2026hpw} for finding exact beyond general-relativity rotating black hole solutions using the formalism in \cite{Li:2026rkf}. 

\clearpage

\setcounter{equation}{0}
\setcounter{section}{0}
\setcounter{subsection}{0}
\setcounter{table}{0}
\setcounter{figure}{0}
\renewcommand{\theequation}{S\arabic{equation}}
\renewcommand{\thesection}{S\Roman{section}}
\renewcommand{\thetable}{S\arabic{table}}
\renewcommand{\thefigure}{S\arabic{figure}}

\section*{Supplementary Material}

%---------------------------------------------------------------------
\section{GHP Formalism}

Here, we briefly summarize the GHP formalism \cite{Geroch:1973am}. 
We adopt the mostly-positive metric convention $(-,+,+,+)$ and the null tetrad $\{l^{\mu}, n^{\mu}, m^{\mu}, \bar{m}^{\mu}\}$ satisfies $l^{\mu}n_{\mu} = -1$, $m^{\mu}\bar{m}_{\mu} = 1$, and all other inner products are zero. 
The metric can be written as 
\begin{equation}
    g_{\mu\nu} = -2l_{(\mu}n_{\nu)} + 2m_{(\mu}\bar{m}_{\nu)}.
\end{equation}
In the GHP formalism, there are three sets of geometric equations that relate the spin coefficients $\{\kappa, \sigma, \rho, \tau\}$, GHP derivatives $\{\thorn, \thorn', \eth, \eth'\}$, the Ricci scalars $\{\Phi_{ij}, \Lambda\}$, and the Weyl scalars $\{\Psi_i\}$. 
For their definitions, see Sec.~4.1 of \cite{Pound:2021qin}. 
First, the GHP derivatives satisfy the following commutation relations: \cite{Geroch:1973am, Bargueno:2023lal}
\begin{align}
    [\thorn', \thorn]f &= [(\bar{\tau}' - \tau)\eth' + (\tau' - \bar{\tau})\eth + p(\kappa\kappa' - \tau'\tau + \Psi_2 + \Phi_{11} - \Lambda) + q(\bar{\kappa}\bar{\kappa}' - \bar{\tau}'\bar{\tau} + \bar{\Psi}_2 + \Phi_{11} - \Lambda)]f, \label{eq:ThornThornCommutator} \\
    [\thorn', \eth']f &= [\bar{\rho}'\eth' + \sigma'\eth - \bar{\tau}\thorn' - \kappa'\thorn + p(\rho\kappa' - \tau\sigma' + \Psi_3) + q(\bar{\sigma}\bar{\kappa}' - \bar{\rho}'\bar{\tau} + \Phi_{21})]f, \label{eq:ThornEthCommutator} \\
    [\eth', \eth]f &= [(\bar{\rho} - \rho)\thorn' + (\rho' - \bar{\rho}')\thorn - p(\rho'\rho - \sigma'\sigma + \Psi_2 - \Phi_{11} - \Lambda) + q(\bar{\rho}'\bar{\rho} - \bar{\sigma}'\bar{\sigma} + \bar{\Psi}_2 - \Phi_{11} - \Lambda)]f, \label{eq:EthEthCommutator} 
\end{align}
and the prime, conjugate, and prime conjugate of these equations. 
The integers $p, q$ represent the GHP-weight of the function $f \overset{\circ}{=} \{p, q\}.$
The Ricci identities -- projections of the Riemann tensor onto tetrads -- are
\begin{align}
    \thorn'\rho' - \eth\kappa' &= \rho'^2 + \sigma'\bar{\sigma}' - \bar{\kappa}'\tau' - \tau\kappa' + \Phi_{22}, \label{eq:ThornRho} \\
    \thorn'\sigma' - \eth'\kappa' &= (\rho' + \bar{\rho}')\sigma' - (\tau' + \bar{\tau})\kappa' + \Psi_4 \\
    \thorn'\tau' - \thorn\kappa' &= (\tau' - \bar{\tau})\rho' + (\bar{\tau}' - \tau)\sigma' + \Psi_3 + \Phi_{21} \label{eq:ThornTau} \\
    \eth'\rho' - \eth\sigma' &= (\rho' - \bar{\rho}')\tau' + (\bar{\rho} - \rho)\kappa' - \Psi_3 + \Phi_{21} \label{eq:EthRho} \\
    \eth'\tau' - \thorn\sigma' &= -\rho\sigma' - \bar{\sigma}\rho' + \tau'^2 + \kappa'\bar{\kappa} + \Phi_{20}, \label{eq:EthTau} \\
    \thorn\rho' - \eth\tau' &= \rho'\bar{\rho} + \sigma'\sigma - \tau'\bar{\tau}' - \kappa'\kappa - \Psi_2 - 2\Lambda, \label{eq:ThornRhoEthTau}
\end{align}
and the prime, conjugate, and prime conjugate of these equations. 
The GHP Bianchi identities are substantially longer and are not explicitly used in our analysis, so we shall direct readers to \cite{Geroch:1973am} for details. \\

To compute the linearized spin coefficients in terms of the metric perturbations $h_{ab}$, following \cite{Loutrel:2020wbw, Li:2026rkf, Spiers:2023cip}, we choose the tetrad perturbation as
\begin{equation}\label{eq:tetrad_pert}
    l^{\mu(1)} = \frac{1}{2}h_{ll}n^{\mu}, \quad n^{\mu(1)} = \frac{1}{2}h_{nn}l^{\mu} + h_{ln}n^{\mu}, \quad m^{\mu(1)} = h_{nm}l^{\mu} + h_{lm}n^{\mu} - \frac{1}{2}h_{m\bar{m}}m^{\mu} - \frac{1}{2}h_{mm}\bar{m}^{\mu}.
\end{equation}
Expanding both sides of the commutation relations and comparing like terms, we have
\begin{align}
    \kappa'^{(1)} &= \frac{1}{2}(\eth' + \tau' - \bar{\tau})h_{nn} - (\thorn' - \bar{\rho}')h_{n\bar{m}} \label{eq:kappa1} \\
    \sigma'^{(1)} &= -\frac{1}{2}(\thorn' + \rho' - \bar{\rho}')h_{\bar{m}\bar{m}} + (\tau' - \bar{\tau})h_{n\bar{m}}\label{eq:sigma1}  \\
    \rho'^{(1)} &= \frac{1}{2}\rho h_{nn} - \frac{1}{2}(\rho' - \bar{\rho}')h_{ln} - \frac{1}{2}(\eth - \tau)h_{n\bar{m}} + \frac{1}{2}(\eth' + 2\tau' - \bar{\tau})h_{nm} - \frac{1}{2}(\thorn' - \rho' + \bar{\rho}')h_{m\bar{m}} \label{eq:rho1} \\
    \tau'^{(1)} &= \frac{1}{2}(\eth' + \tau' - \bar{\tau})h_{ln} - \frac{1}{2}(\thorn - \rho)h_{n\bar{m}} - \frac{1}{2}(\thorn' - \bar{\rho}')h_{l\bar{m}} - \frac{1}{2}\tau h_{\bar{m}\bar{m}} - \frac{1}{2}\bar{\tau}h_{m\bar{m}}, \label{eq:tau1}
\end{align}
and the prime, conjugate, and prime conjugate of these equations. 

%---------------------------------------------------------------------
\section{Primed Held's integration technique}\label{sec:Held}

In the original Held's formalism \cite{Held:1974for, Held:1976kil}, all expressions are written in terms of $\rho$ or Held scalars that are annihilated by $\thorn$, denoted with a degree mark $^\circ$, so that GHP derivatives can be easily applied to various quantities. 
In our case, since we are using the ORG and we primarily use $\rho'$ rather than $\rho$, we would like the primed variant of this, so that  Held scalars are annihilated by $\thorn'$. 
Taking the GHP prime of the original version given in \cite{Held:1974for, Held:1976kil, Hollands:2024iqp}, we define the following:
\begin{align}
    \rho &= \frac{1}{2}(\rho' + \bar{\rho}')\rho^{\circ} + \frac{1}{2}(\rho' - \bar{\rho}')\bar{\rho}^{\circ} - \frac{1}{2} \rho'^2 \Psi'^{\circ} - \frac{1}{2}\rho'\bar{\rho}' \bar{\Psi}'^{\circ} - \rho'^2\bar{\rho}'\tau'^{\circ}\bar{\tau}'^{\circ} \label{eq:rhocirc} \\
    \tau &= -\bar{\tau}'^{\circ} \rho'^2 \label{eq:taubarcirc} \\
    \tau' &= \tau'^{\circ} \rho'\bar{\rho}' \label{eq:taucirc} \\
    \Psi_2 &= \Psi'^{\circ} \rho'^3 \label{eq:Psicirc} \\
    \Omega'^{\circ} &= \frac{1}{\bar{\rho}'} - \frac{1}{\rho'}. \label{eq:Omegacirc}
\end{align}
We also define tilde GHP derivatives 
\begin{align}
    \thorn &= \tilde{\thorn} + \rho'\bar{\rho}'\bar{\tau}'^{\circ} \tilde{\eth}' + \rho'\bar{\rho}'\tau'^{\circ}\tilde{\eth} + \rho'\bar{\rho}'\tau'^{\circ}\bar{\tau}'^{\circ}(p\rho' + q\bar{\rho}') + \frac{1}{2}\left(p\Psi'^{\circ}\rho'^2 + q\bar{\Psi}'^{\circ}\bar{\rho}'^2\right) \\
    \eth' &= \bar{\rho}' \tilde{\eth}' + q\bar{\rho}'^2\tau'^{\circ} \\
    \eth &= \rho' \tilde{\eth} + p\rho'^2\bar{\tau}'^{\circ}.
\end{align}
These tilde GHP derivatives, when acting on Held scalars, yield Held scalars. 
Specifically, Table~\ref{tab:DerivativesOfHeldQuantities} summarizes tilde derivatives acting on $\rho'$ and Held scalars $\rho^{\circ}$, $\tau'^{\circ}$, $\Psi'^{\circ}$ and $\Omega'^{\circ}$. 
Furthermore, these tilde GHP derivatives satisfy the following commutation relations:
\begin{align}
    [\thorn', \tilde{\thorn}] &= \left(\rho'\bar{\rho}'(\rho' + \bar{\rho}')\tau'^{\circ}\bar{\tau}'^{\circ} + \frac{1}{2}\rho'^2\Psi'^{\circ} + \frac{1}{2}\bar{\rho}'^2\bar{\Psi}'^{\circ}\right) \thorn', \\
    [\thorn', \tilde{\eth}] &= 0, \\
    [\tilde{\eth}', \tilde{\eth}] &= \left(-\frac{\rho^{\circ}}{\bar{\rho}'} + \frac{\bar{\rho}^{\circ}}{\rho'} + \frac{1}{2}\rho'\Omega'^{\circ}\Psi'^{\circ} + \frac{1}{2}\bar{\rho}'\Omega'^{\circ}\bar{\Psi}'^{\circ} + \rho'\bar{\rho}'\Omega'^{\circ}\tau'^{\circ}\bar{\tau}'^{\circ}\right)\thorn' + \Omega'^{\circ}\tilde{\thorn} - p\rho^{\circ} + q\bar{\rho}^{\circ}, \\
    [\tilde{\thorn},\tilde{\eth}] &= \left(\rho'\rho^{\circ}\bar{\tau}'^{\circ} - \bar{\rho}'\bar{\rho}^{\circ}\bar{\tau}'^{\circ} - \frac{1}{2}\rho'^2\bar{\rho}'\bar{\tau}'^{\circ}\Omega'^{\circ}\Psi'^{\circ} - \frac{1}{2}\rho'\bar{\rho}'^2\bar{\tau}'^{\circ}\Omega'^{\circ}\bar{\Psi}'^{\circ} - \rho'^2\bar{\rho}'^2\Omega'^{\circ}\tau'^{\circ}\bar{\tau}'^{\circ 2}\right)\thorn'.
\end{align}

\begin{table}[t]
\centering
\renewcommand{\arraystretch}{1.5}
\setlength{\tabcolsep}{10pt}
\begin{tabular}{c|ccc}
& $\tilde{\thorn}$ & $\tilde{\eth}$ & $\tilde{\eth}'$ \\ \hline
\rule{0pt}{1.8em}
$\rho'$ & 
$\rho'^2 \rho^{\circ} - \half \rho'^2 (\rho'\Psi'^{\circ} + \bar{\rho}'\bar{\Psi}'^{\circ}) - \rho'^3 \bar{\rho}' \tau'^\circ \bar{\tau}'^\circ$ &
$-\bar{\tau}'^\circ \rho'^2$ &
$\tau'^\circ \rho'^2$ \\
$\Omega'^\circ$ & $0$ & $-2\bar{\tau}'^\circ$ & $2\tau'^\circ$\\
$\rho^{\circ}$ & $0$ & $0$ & $0$ \\
$\tau'^\circ$ & $0$ & $\half(\rho^{\circ} + \bar{\rho}^{\circ})\Omega'^\circ + \half(\Psi'^\circ - \bar{\Psi}'^\circ)$ & $0$\\
$\Psi'^\circ$ & $0$ & $0$ & $0$ \\
\end{tabular}
\caption{Tilde GHP derivative of Held quantities in type D spacetimes.}
\label{tab:DerivativesOfHeldQuantities}
\end{table}

%---------------------------------------------------------------------
\section{Teukolsky and Einstein Operators in GHP form}\label{sec:Operator}
\noindent
Taken from \cite{Pound:2021qin}, $\Psi_4$ can be expressed in terms of $h_{ab}$ as follows:
\begin{align}\label{eq:T4}
    \Psi_4 = \mathcal{T}_4 h &= -\frac{1}{2}\Big[(\eth'-\bar{\tau})(\eth'-\bar{\tau})h_{nn} + (\thorn'-\bar{\rho}')(\thorn'-\bar{\rho}')h_{\bar m\bar m} \\
    &\qquad\qquad - \big((\thorn'-\bar{\rho}')(\eth'-2\bar{\tau})+(\eth'-\bar{\tau})(\thorn'-2\bar{\rho}')\big)h_{(n\bar m)}\Big],
\end{align}
and the adjoint is 
\begin{align}\label{eq:T4dagger}
    ({\cal T}_4^{\dagger}\eta)_{ab} &= -\frac{1}{2}\Big[n_an_b(\eth' - \tau')(\eth' - \tau') + \bar{m}_a\bar{m}_b(\thorn' - \rho')(\thorn' - \rho') \\
    &\qquad\qquad - n_{(a}\bar{m}_{b)}((\eth' - \tau' + \bar{\tau})(\thorn' - \rho') + (\thorn' - \rho' + \bar{\rho}')(\eth' - \tau'))\Big]\eta.
\end{align}
The Teukolsky equation is
\begin{equation}
    {\cal O}_4\Psi_4 = {}_{-2}T, 
\end{equation}
where ${\cal O}_4$ is the Teukolsky operator
\begin{equation}
    {\cal O}_4 = (\thorn' - 4\rho' - \bar{\rho}')(\thorn - \rho) - (\eth' - 4\tau' - \bar{\tau})(\eth - \tau) - 3\Psi_2,
\end{equation}
and ${}_{-2}T$ is the source of the Teukolsky equation 
\begin{align}\label{eq:S4}
    {}_{-2}T = {\cal S}_4T &= \frac{1}{2}(\eth' - \bar{\tau} - 4\tau')\left[(\thorn' - 2\bar{\rho}')T_{(n\bar{m})} - (\eth' - \bar{\tau})T_{nn}\right] \nonumber \\
    &\qquad\qquad + \frac{1}{2}(\thorn' - 4\rho' - \bar{\rho}')\left[(\eth' - 2\bar{\tau})T_{(n\bar{m})} - (\thorn' - \bar{\rho}')T_{\bar{m}\bar{m}}\right],
\end{align}
with the adjoint 
\begin{align}\label{eq:S4dagger}
    ({\cal S}_4^{\dagger}\Phi)_{ab} &= -\frac{1}{2}\Big[n_an_b(\eth' - \tau')(\eth' + 3\tau') + \bar{m}_a\bar{m}_b(\thorn' - \rho')(\thorn' + 3\rho') \nonumber \\
    &\qquad\qquad - n_{(a}\bar{m}_{b)}((\eth' - \tau' + \bar{\tau})(\thorn' + 3\rho') + (\thorn' - \rho' + \bar{\rho}')(\eth' + 3\tau'))\Big]\Phi.
\end{align}
These operators form the Wald identity ${\cal S}_4{\cal E} = {\cal O}_4{\cal T}_4$, where ${\cal E}$ is the self-adjoint Einstein operator given in the Supplementary Material [Sec.~\ref{sec:Operator}]. 

The divergence of a symmetric $(0,2)$-tensor $F_{ab}$ is \cite{Toomani:2021jlo} 
\begin{align}\label{eq:BianchiFull}
    \nabla^b F_{ab} = -l_a &\left[(\thorn' - \rho' - \bar{\rho}')F_{ln} + (\thorn - \rho - \bar{\rho})F_{nn} - (\eth' - 2\tau' - \bar{\tau})F_{nm} - (\eth - \tau - 2\bar{\tau}')F_{n\bar{m}} - (\rho' + \bar{\rho}')F_{m\bar{m}}\right] \nonumber \\
    - n_a &\left[(\thorn' - \rho' - \bar{\rho}')F_{ll} + (\thorn - \rho - \bar{\rho})F_{ln} - (\eth' - \tau' - 2\bar{\tau})F_{lm} - (\eth - 2\tau - \bar{\tau}')F_{l\bar{m}} - (\rho + \bar{\rho})F_{m\bar{m}}\right] \nonumber \\
    + m_a &\left[(\tau' + \bar{\tau})F_{ln} + (\thorn' - 2\rho' - \bar{\rho}')F_{l\bar{m}} + (\thorn - \rho - 2\bar{\rho})F_{n\bar{m}} - (\eth' - \tau' - \bar{\tau})F_{m\bar{m}} - (\eth - \tau - \bar{\tau}')F_{\bar{m}\bar{m}}\right] \nonumber \\
    + \bar{m}_a &\left[(\tau + \bar{\tau}')F_{ln} + (\thorn' - \rho' - 2\bar{\rho}')F_{lm} + (\thorn - 2\rho - \bar{\rho})F_{nm} - (\eth' - \tau' - \bar{\tau})F_{mm} - (\eth - \tau - \bar{\tau}')F_{m\bar{m}}\right].
\end{align}

The linearized Einstein operator in a Petrov type II spacetime was given in \cite{Price:2007dev, Green:2019nam}, and specialized to Petrov type D in \cite{Hollands:2024iqp}. 
These Einstein operators can be derived from taking suitable linear combinations of the linearized Ricci identities and commutation relations. 
Here, we quote the final expression stated in \cite{Hollands:2024iqp, Wardell:2024yoi}.
\begin{equation}
\begin{split}
-({\cal E}h)_{ll} &= \Big[(\eth'-\tau')(\eth-\bar{\tau}') + \rho(\thorn'+\rho'-\bar{\rho}') - (\thorn-\rho)\rho' + \Psi_2\Big]h_{ll} \\
&\pheq + \Big[-(\rho+\bar{\rho})(\thorn+\rho+\bar{\rho}) + 4\rho\bar{\rho}\Big]h_{ln} \\
&\pheq + \Big[-(\thorn-3\bar{\rho})(\eth'-\tau'+\bar{\tau}) + \bar{\tau}\thorn-\bar{\rho}\eth'\Big]h_{lm} \\
&\pheq + \Big[-(\thorn-3\rho)(\eth+\tau-\bar{\tau}') + \tau\thorn-\rho\eth\Big]h_{l\bar{m}} \\
&\pheq + \Big[\thorn(\thorn-\rho-\bar{\rho})+2\rho\bar{\rho}\Big]h_{m\bar{m}},
\end{split}
\label{eq:Ell_full}
\end{equation}

\begin{equation}
\begin{split}
-({\cal E}h)_{nn} 
&= \pheq  \Big[(\eth'-\bar{\tau})(\eth-\tau) +\bar{\rho}'(\thorn-\rho+\bar{\rho}) -
(\thorn'-\bar{\rho}')\bar{\rho} + {\bar\Psi}_2\Big]h_{nn}\\
&\pheq + \Big[-(\rho'+\bar{\rho}')(\thorn'+\rho'+\bar{\rho}') + 4\rho'\bar{\rho}'
 \Big]h_{ln}\\
&\pheq + \Big[-(\thorn'-3\rho')(\eth'+\tau'-\bar{\tau}) + \tau'\thorn' - \rho'\eth'\Big]h_{nm}\\
&\pheq + \Big[-(\thorn'-3\bar{\rho}')(\eth+\bar{\tau}'-\tau) + \bar{\tau}'\thorn' - \bar{\rho}'\eth\Big]h_{n\bar{m}}\\
&\pheq + \Big[\thorn'(\thorn'-\rho'-\bar{\rho}') + 2\rho'\bar{\rho}'\Big]h_{m\bar{m}},
\end{split}
\label{eq:Enn_full}
\end{equation}

\begin{equation}
\begin{split}
-({\cal E}h)_{ln} &= \half\Big[\rho'(\thorn'-\rho') + \bar{\rho}'(\thorn'-\bar{\rho}')\Big]h_{ll}\\
&\pheq + \half\Big[\rho(\thorn-\rho) + \bar{\rho}(\thorn-\bar{\rho})\Big]h_{nn}\\
&\pheq + \half\Big[-(\eth'+\tau'+\bar{\tau})(\eth-\tau-\bar{\tau}') -
(\eth'\eth+3\tau\tau'+3\bar{\tau}\bar{\tau}') + 2(\bar{\tau}+\tau')\eth\\
&\pheq\pheq+ (\thorn-2\bar{\rho})\rho' +(\thorn'-2\rho')\bar{\rho} - \bar{\rho}'(\thorn+\rho) -
\rho(\thorn'+\bar{\rho}') -\Psi_2 -{\bar\Psi}_2\Big]h_{ln}\\
&\pheq + \half\Big[(\thorn'-2\bar{\rho}')(\eth'-\tau') + \bar{\tau}(\thorn'+\rho'+\bar{\rho}') - \tau'(\thorn'-\rho') - (2\eth'-\bar{\tau})\bar{\rho}'\Big]h_{lm}\\
&\pheq + \half\Big[(\thorn'-2\rho')(\eth-\bar{\tau}') + \tau(\thorn'+\bar{\rho}'+\rho') - \bar{\tau}'(\thorn'-\bar{\rho}') - (2\eth-\tau)\rho'\Big]h_{l\bar{m}}\\
&\pheq + \half\Big[(\thorn-2\rho)(\eth'-\bar{\tau}) + (\tau'+\bar{\tau})(\thorn+\bar{\rho})
-2(\eth'-\tau')\rho-2\bar{\tau}\thorn\Big]h_{nm}\\
&\pheq + \half\Big[(\thorn-2\bar{\rho})(\eth-\tau) + (\bar{\tau}'+\tau)(\thorn+\rho)
-2(\eth-\bar{\tau}')\bar{\rho}-2\tau\thorn\Big]h_{n\bar{m}}\\
&\pheq + \half\Big[-(\eth'-\bar{\tau})(\eth'-\tau') +
\bar{\tau}(\bar{\tau}-\tau')\Big]h_{mm}\\
&\pheq + \half\Big[-(\eth-\tau)(\eth-\bar{\tau}') +
\tau(\tau-\bar{\tau}')\Big]h_{\bar{m}\bar{m}}\\
&\pheq + \half\Big[(\eth'+\tau'-\bar{\tau})(\eth-\tau+\bar{\tau}') +
(\eth'\eth-\tau\tau'-\bar{\tau}\bar{\tau}'+\tau\bar{\tau}) - (\Psi_2+{\bar\Psi}_2) + (\thorn'-2\rho')\bar{\rho} + (\thorn-2\bar{\rho})\rho' \\
&\pheq\pheq +\rho(3\thorn'-2\bar{\rho}') + \bar{\rho}'(3\thorn-2\rho) - 2\thorn'\thorn + 2\rho\bar{\rho}' +2\eth'(\tau)-\tau\bar{\tau}\Big]h_{m\bar{m}},
\end{split}
\label{eq:Eln_full}
\end{equation}

\begin{equation}
\begin{split}
-({\cal E}h)_{lm} &= \half\Big[(\thorn'-\rho')(\eth-\bar{\tau}')
+(\eth-\tau-2\bar{\tau}')\bar{\rho}' -(\eth-\tau)\rho' +\tau(\thorn'+\rho')\Big]h_{ll}\\
&\pheq +\half\Big[-(\thorn-\rho+\bar{\rho})(\eth+\tau-\bar{\tau}') -
(\eth-3\tau+\bar{\tau}')\bar{\rho} - 2\rho\bar{\tau}'\Big]h_{ln}\\
&\pheq +\half\Big[-(\thorn'+\bar{\rho}')(\thorn-2\bar{\rho}) + \rho(\thorn'+2\rho'-2\bar{\rho}') -
4\rho'\bar{\rho} +3\Psi_2\\
&\pheq\pheq + (\eth'+\bar{\tau})(\eth-2\bar{\tau}') - \tau(\eth'+\tau'-2\bar{\tau})
-\tau'(\tau-4\bar{\tau}')\Big]h_{lm}\\
&\pheq + \half\Big[-\eth(\eth-2\tau)  -
2\bar{\tau}'(\tau-\bar{\tau}')\Big]h_{l\bar{m}}\\
&\pheq + \half\Big[\thorn(\thorn-2\rho) + 2\bar{\rho}(\rho-\bar{\rho})\Big]h_{nm}\\
&\pheq + \half\Big[-(\thorn-\bar{\rho})(\eth'-\tau'+\bar{\tau}) + 2\bar{\tau}\bar{\rho}\Big]h_{mm}\\
&\pheq + \half\Big[(\thorn+\rho-\bar{\rho})(\eth+\bar{\tau}'-\tau) + 2\bar{\tau}'(\thorn-2\rho) -
(\eth-\tau-\bar{\tau}')\bar{\rho} +2\rho\tau\Big]h_{m\bar{m}},
\end{split}
\label{eq:Elm_full}
\end{equation}

\begin{equation}
\begin{split}
-({\cal E}h)_{n\bar{m}} &= 
\pheq  \half\Big[(\thorn-\rho+\bar{\rho})(\eth'-\bar{\tau}) - (\eth'-2\tau'+\bar{\tau})\rho +
\tau'(\thorn-\bar{\rho})\Big]h_{nn}\\
&\pheq + \half\Big[-(\thorn'-\rho'+\bar{\rho}')(\eth'+\tau'-\bar{\tau}) -
(\eth'-3\tau'+\bar{\tau})\bar{\rho}' -2\rho'\bar{\tau}\Big]h_{ln}\\
&\pheq +\half\Big[\thorn'(\thorn'-2\rho') +2\bar{\rho}'(\rho'-\bar{\rho}')\Big]h_{l\bar{m}}\\
&\pheq +\half\Big[-\eth'(\eth'-2\tau') -
2\bar{\tau}(\tau'-\bar{\tau})\Big]h_{nm}\\
&\pheq +\half\Big[-(\thorn+\bar{\rho})(\thorn'-2\bar{\rho}') + \rho'(\thorn+2\rho-2\bar{\rho}) -
4\rho\bar{\rho}' + 3\Psi_2\\
&\pheq\pheq + (\eth+\bar{\tau}')(\eth'-2\bar{\tau}) - \tau'(\eth+\tau-2\bar{\tau}') -
\tau(\tau'-4\bar{\tau})\Big]h_{n\bar{m}}\\
&\pheq +\half\Big[-(\thorn'-\bar{\rho}')(\eth-\tau+\bar{\tau}') +2\bar{\tau}'\bar{\rho}'\Big]h_{\bar{m}\bar{m}}\\
&\pheq +\half\Big[(\thorn'+\rho'-\bar{\rho}')(\eth'-\tau'+\bar{\tau}) + 2\bar{\tau}(\thorn'-2\rho')
-(\eth'-\tau'-\bar{\tau})\bar{\rho}'+2\rho'\tau'\Big]h_{m\bar{m}} \\
\end{split}
\label{eq:Enmb_full}
\end{equation}

\begin{equation}
\begin{split}
-({\cal E}h)_{mm} &= 
\pheq \Big[-\eth(\eth-\tau-\bar{\tau}') -2\tau\bar{\tau}'\Big]h_{ln}\\
&\pheq +\Big[(\thorn'-\rho')(\eth-\bar{\tau}') - (\eth-\tau-\bar{\tau}')\rho' +
\tau(\thorn'+\rho'-\bar{\rho}') -\bar{\tau}'(\thorn'+\bar{\rho}')\Big]h_{lm}\\
&\pheq +\Big[(\thorn-\bar{\rho})(\eth-\tau) - (\eth-\tau-\bar{\tau}')\bar{\rho} -\tau(\thorn+\rho)
+ \bar{\tau}'(\thorn-\rho+\bar{\rho})\Big]h_{nm}\\
&\pheq +\Big[-(\thorn'-\rho')(\thorn-\bar{\rho}) + (\eth-\tau)\tau' -
\tau(\eth'+\tau'-\bar{\tau}) + \Psi_2\Big]h_{mm}\\
&\pheq +\Big[(\tau+\bar{\tau}')\eth +
(\tau-\bar{\tau}')^2\Big]h_{m\bar{m}},
\end{split}
\label{eq:Emm_full}
\end{equation}

\begin{equation}
\begin{split}
-({\cal E}h)_{m\bar{m}} &= \half\Big[\thorn'(\thorn'-\rho'-\bar{\rho}') + 2\rho'\bar{\rho}' \Big]h_{ll}\\
&\pheq +\half\Big[\thorn(\thorn-\rho-\bar{\rho}) + 2\rho\bar{\rho}\Big]h_{nn}\\
&\pheq +\half\Big[-(\thorn'+\rho'-\bar{\rho}')(\thorn-\rho+\bar{\rho}) -\thorn'(\thorn+\rho)
+\rho(\thorn'+\rho'-\bar{\rho}') -{\bar\Psi}_2 + (\eth'-\bar{\tau})(\eth-\tau-\bar{\tau}') \\
&\pheq\pheq + \eth'\eth - (\eth-2\bar{\tau}')\tau' - \bar{\tau}(2\eth+\bar{\tau}') - 2\tau(\eth'-\bar{\tau})+2\tau'\bar{\tau}'+\bar{\rho}\bar{\rho}'\Big]h_{ln}\\
&\pheq +\half\Big[-(\thorn'-2\rho')(\eth'-2\bar{\tau}) + \bar{\tau}(\thorn'+2\rho'-2\bar{\rho}')-2\tau'\bar{\rho}'\Big]h_{lm}\\
&\pheq +\half\Big[-(\thorn'-2\bar{\rho}')(\eth-2\tau) + \tau(\thorn'+2\bar{\rho}'-2\rho') -2\bar{\tau}'\rho'\Big]h_{l\bar{m}}\\
&\pheq +\half\Big[-(\thorn-2\bar{\rho})(\eth'-2\tau') + \tau'(\thorn-2\rho-2\bar{\rho}) -
2\rho\bar{\tau}+4\tau'\bar{\rho}\Big]h_{nm}\\
&\pheq +\half\Big[-(\thorn-2\rho)(\eth-2\bar{\tau}') + \bar{\tau}'(\thorn-2\bar{\rho}-2\rho) -
2\bar{\rho}\tau+4\bar{\tau}'\rho\Big]h_{n\bar{m}}\\
&\pheq
+\half\Big[-\bar{\tau}(\eth'-\bar{\tau})-\tau'(\eth'-\tau')
\Big]h_{mm}\\
&\pheq +\half\Big[-\tau(\eth-\tau)-\bar{\tau}'(\eth-\bar{\tau}')
\Big]h_{\bar{m}\bar{m}}\\
&\pheq + \half\Big[2\thorn'\thorn -(\thorn'-\bar{\rho}')\bar{\rho} - (\thorn-\rho)\rho'
-\rho(\thorn'-\rho'+\bar{\rho}') - \bar{\rho}'(\thorn+\rho-\bar{\rho})\\
&\pheq\pheq -(\eth'-2\tau')\bar{\tau}' + \tau(\eth'+2\bar{\tau}) -
\tau'(\eth-\bar{\tau}') + \bar{\tau}(\eth+\tau) -\eth'(\tau) - \Psi_2 - {\bar\Psi}_2\Big]h_{m\bar{m}}.
\end{split}
\label{eq:Emmb_full}
\end{equation}

\section{Two Examples}
\subsection{Static Deformation: Hernquist Dark Matter Halo}

We apply the Einstein reconstruction scheme to reconstruct the metric deformation induced by a Hernquist dark matter halo profile in a Schwarzschild background. We use this example (instead of a dark matter halo in a Kerr background) because, for this specific case, an exact metric deformation is known in closed form in Schwarzschild coordinates~\cite{Cardoso:2021wlq}, namely
\begin{align}\label{eq:DM_metric}
    ds^2 = -g(R) dt^2 + \frac{dR^2}{1 - 2m(R)/R} + R^2 d\Omega^2, 
\end{align}
where
\begin{gather}
    g(R) = f(R) e^{\Upsilon}, \quad m(R) = M + \frac{M_c R^2}{(R + a_0)^2} f(R)^2, \quad f(R) = 1 - \frac{2M}{R}, \\ 
    \Upsilon = -\pi\sqrt{\frac{M_c}{b_0}} + 2\sqrt{\frac{M_c}{b_0}} \arctan\frac{R + a_0 - M_c}{\sqrt{M_c b_0}}, \quad b_0 = 2a_0 - M_c + 4M.
\end{gather}
The quantity $a_0$ represents the typical length scale of the halo, and $M_c$ and $M$ are the masses of the halo and the black hole, respectively.
Reference~\cite{Cardoso:2021wlq} has shown that this metric satisfies the nonlinear Einstein equation with stress-energy tensor $T^{\mu}{}_{\nu} = \mathrm{diag}(-\rho, 0, P_t, P_t)$,  
\begin{align}
    \rho(R) = \frac{4M_c(a_0 + 2M)f(R)}{R(R+a_0)^3}, \qquad 2P_t(R) = \frac{M \rho(R)}{R - 2M}.
\end{align}
Having an exact solution is useful because we can compare it to our Einstein reconstruction solution (which we derive below) and check for the correctness of the latter.

Let us then assume that $M_c = \epsilon M \ll M$ with $\epsilon \ll 1$ to Einstein-reconstruct the metric perturbation and show that the result recovers the above metric in the perturbative regime.
First, we linearize the stress-energy tensor and project it onto the tetrads defined in Eq.~\eqref{eq:tetrad}; the non-zero components are\footnote{Notice that the asymptotics of $T_{ab}$ satisfy the requirements listed in Sec.~\ref{sec:Theorem2Proof}, except $T_{nn} = O(r^{-4})$. Nevertheless, we find that no homogeneous solution in Eq.~\eqref{eq:CompleteSolution} is required.}
\begin{align}
    T_{nn} = 2f(r)T_{ln} = 4f(r)^2T_{ll} = \frac{2r}{M}T_{m\bar{m}} =\frac{4\epsilon M(a_0 + 2M)}{r(r+a_0)^3}. 
\end{align}
Since $T_{n\bar{m}} = T_{\bar{m}\bar{m}} = 0$, and $T_{nn}$ is spherically symmetric, ${}_{-2}T = 0$. 
This implies $\Psi_4 = 0$, as the homogeneous solutions of the Teukolsky equation are not compatible with the boundary conditions. 

Next, we reconstruct the metric deformation hierarchically using the particular solutions of Sec.~\ref{sec:ParticularSolutions}. 
Using the fact that 
\begin{equation}
    \int d\lambda = \Big(\frac{\Gamma}{\bar{\Gamma}}\Big)^{-s}\int_r^\infty \,dr' \Big(\frac{\Gamma'}{\bar{\Gamma}'}\Big)^s 
\end{equation}
the spin-0 $h_{m\bar{m}}^p$ ($s=0$) component is 
\begin{equation}
    h_{m\bar{m}}^p = \int_r^{\infty} r' T_{nn} \,dr' + \frac{1}{r} \int_r^{\infty} r^{\prime\,2} T_{nn} \,dr' = -\frac{2\epsilon M(a_0 + 2M)}{r(r + a_0)}.
\end{equation}
Since $\Psi_4 = 0$, the spin-2 component $h_{\bar{m}\bar{m}}^p = 0$, while the spin-1 component $h_{l\bar{m}}^p$ also vanishes since $\tilde{T}_{n\bar{m}} = 0$. 
Lastly, the spin-0 component $h_{ll}^p$ is 
\begin{equation}
    h_{ll}^p = -\frac{1}{r} \int_r^{\infty} r^{\prime\,2} \tilde{T}_{ln} = -\frac{1}{r} \int_r^{\infty} \frac{2\epsilon M\,(a_0 + 2M)(r^{\prime\,2} - 2Mr' - a_0 M)}{r^{\prime\,2}(r' + a_0)^2} \,dr' = -\frac{2\epsilon M (a_0 + 2M) (r - M)}{r^2 (r + a_0)}.
\end{equation}
Since $T_{ab}$ decays sufficiently fast, we find that no homogeneous solution in Eq.~\eqref{eq:CompleteSolution} is required.
However, we need to include an additional mass shift $2\epsilon M/r$ to $h_{ll}$ to make sure that the ADM mass is the same as in Eq.~\eqref{eq:DM_metric} to linear order in $\epsilon$. Therefore, 
\begin{equation}
    ds^2 = \left(-f(r) + \frac{2\epsilon M}{r} + h_{ll}\right) dv^2 + 2dvdr + r^2 (1 + h_{m\bar{m}}) d\Omega^2.
\end{equation}

To cast this result in the same form as Eq.~\eqref{eq:DM_metric}, we first convert the radius $r$ into the areal radius $R$, 
\begin{equation}
    R^2 = r^2 (1 + h_{m\bar{m}}) \quad \Rightarrow \quad r = R + \frac{\epsilon M (a_0 + 2M)}{R + a_0} + O(\epsilon^2). 
\end{equation}
We also need to transform the metric to Schwarzschild coordinates via 
\begin{equation}
    dv = dt + \left(f(r) - \frac{2\epsilon M}{r} - h_{ll}\right)^{-1} dr. 
\end{equation}
Applying both coordinate transformations, we then have 
\begin{align}
    ds^2 = -f(R) dt^2 + \frac{dR^2}{f(R)} + R^2 d\Omega^2 + \epsilon \left[\frac{2M f(R)}{R + a_0}dt^2 + \frac{2MR}{(R + a_0)^2} dR^2\right].
\end{align}
which coincides exactly with Eq.~\eqref{eq:DM_metric}, after linearizing in $\epsilon$.

\subsection{Dynamical Perturbation: Linearized Vaidya Spacetime}

The Vaidya spacetime models a radiating spacetime, whose metric is 
\begin{equation}\label{eq:Vaidya}
    ds^2 = -f(r) dv^2 + 2dvdr + r^2 d\Omega^2, \qquad f(r) = 1 - \frac{2M(v, \theta, \varphi)}{r}, 
\end{equation}
and the stress-energy tensor associated with it is 
\begin{equation}\label{eq:Vaidya_Source}
    T_{ab} = \frac{2\dot{M}(v, \theta, \varphi)}{r^2} n_a n_b. 
\end{equation}
As is well known, this is an exact solution to the Einstein equations, so it serves as a useful second test of our Einstein reconstruction method. 

Let us then use the the source in Eq.~\eqref{eq:Vaidya_Source} (in a Schwarzschild background and with mass $M_0$) to reconstruct the metric perturbation, and to check whether the reconstructed solution recovers the Vaidya metric to leading order in the perturbative regime.
Let us assume that, at time $v=0$, a small ingoing mass flux $|\dot{M}| \ll 1$ enters the black hole's event horizon. 
Since the only non-zero stress-energy component is $T_{ll}$, the Teukolsky source $_{-2}T = 0$, and all sources in the transport equations are zero. 
Therefore, all particular solutions $h_{ab}^p$ vanish. 
Different from the previous example, this time we need the homogeneous fixing procedure, because the $(l,l)$-component of the Einstein equations 
is not satisfied. Its residual due to the particular solution is 
\begin{equation}
    F_{ll}^p = -T_{ll} = -\frac{2\dot{M}(v, \theta, \varphi)}{r^2}.
\end{equation}
Using Eq.~\eqref{eq:Fll_p}, we find $p_{ll}^{\circ} = -2\dot{M}$ and other $p_{ab}^{\circ} = 0$. 
Thus, Eq.~\eqref{eq:pll} tells us that 
\begin{equation}\label{eq:Vaidya_Flux_Balance}
    d_{ll}^{\circ} = \int_0^v \dot{M}(v', \theta, \varphi) \,dv' \equiv \delta M(v, \theta, \varphi). 
\end{equation}
Therefore, the reconstructed spacetime is
\begin{equation}
    ds^2 = -\left(1 - \frac{2[M_0 + \delta M(v, \theta, \varphi)]}{r}\right) dv^2 + 2dvdr + r^2 d\Omega^2,
\end{equation}
which is the linearized version of the Vaidya spacetime, given in Eq.~\eqref{eq:Vaidya}. 
Note that Eq.~\eqref{eq:Vaidya_Flux_Balance} represents the usual flux balance law for Vaidya spacetime.

\twocolumngrid
\bibliography{ref.bib}

\end{document}